\documentclass[aps,prd,superscriptaddress,onecolumn,showpacs,nofootinbib,preprintnumbers,notitlepage]{revtex4-1}

\usepackage{amsmath,amssymb,hyperref,subfigure,xspace}
\usepackage{mciteplus,color,mathtools,graphicx}
\usepackage{multirow}
\usepackage[english]{babel}
\usepackage{xspace}
\usepackage{slashed}
\newcommand{\ls}{LS\xspace}

\newcommand{\cpt}{CPM\xspace}

\newcommand{\lambdab}{\ensuremath{\lambda_b}}
\newcommand{\lambdapp}{\ensuremath{\lambda_p}}
\newcommand{\lambdapsi}{\ensuremath{\lambda_\psi}}
\newcommand{\Kp}{\ensuremath{K^+}}
\newcommand{\Km}{\ensuremath{K^-}}
\newcommand{\Eb}{\ensuremath{E_b}}
\newcommand{\Epp}{\ensuremath{E_p}}
\newcommand{\Epsi}{\ensuremath{E_\psi}}

\newcommand{\pb}{\ensuremath{p_b}}
\newcommand{\ppp}{\ensuremath{p_p}}
\newcommand{\ppsi}{\ensuremath{p_\psi}}
\newcommand{\pK}{\ensuremath{p_K}}
\newcommand{\pbarpsi}{\ensuremath{\bar{p}_\psi}}
\newcommand{\pbarK}{\ensuremath{\bar{p}_K}}
\newcommand{\mb}{\ensuremath{m_b}}
\newcommand{\mpp}{\ensuremath{m_p}}
\newcommand{\mpsi}{\ensuremath{m_\psi}}
\newcommand{\mK}{\ensuremath{m_K}}
\newcommand{\slashedppsi}{\ensuremath{\slashed{p}_\psi}}
\newcommand{\slashedpb}{\ensuremath{\slashed{p}_b}}
\newcommand{\slashedppp}{\ensuremath{\slashed{p}_p}}
\newcommand{\lambdai}{\ensuremath{\lambda_{b\psi}}}
\newcommand{\lambdaf}{\ensuremath{\lambda_{pK}}}
\newcommand{\mmf}[1]{{\ensuremath{Q^{#1}}}}

\DeclareMathOperator{\re}{Re}

\DeclareMathOperator{\tr}{Tr}
\newcommand{\babar}{BaBar}
\newcommand{\belle}{Belle\xspace}
\newcommand{\lhcb}{LHCb\xspace}

\newcommand{\ie}{{\it i.e.}\xspace}

\newcommand{\simeqzero}[1]{\ensuremath{\xrightarrow[#1]{\hspace*{.5cm}} 0}\xspace}
\newcommand{\helamp}{\ensuremath{\mathcal{A}_{\lambdapp, \lambdab \lambdapsi}}\xspace}
\newcommand{\heliso}{\ensuremath{A_{\lambdapp, \lambdab \lambdapsi}}\xspace}
\newcommand{\hheliso}{\ensuremath{\hat A_{\lambdapp, \lambdab \lambdapsi}}\xspace}
\newcommand{\helisou}{\ensuremath{A_{\lambdapp \lambdapsi, \lambdab}}\xspace}
\newcommand{\hhelisou}{\ensuremath{\hat A_{\lambdapp \lambdapsi, \lambdab }}\xspace}
\newcommand{\lambdap}{\ensuremath{\lambda^\prime}\xspace}
\newcommand{\abs}[1]{\ensuremath{\left|#1\right|}}
\newcommand{\jpsi}{{\psi}}
\newcommand{\Lambdab}{\ensuremath{\Lambda_b}\xspace}
\newcommand{\slashedP}{\ensuremath{\slashed{P}}}
\newcommand{\propagatorP}{\ensuremath{P}}
\newcommand{\ppsibar}{\ensuremath{\bar{p}_\psi}}
\newcommand{\Lambdares}{\ensuremath{\Lambda^*}}
\newcommand{\mev}{\ensuremath{{\mathrm{\,Me\kern -0.1em V}}}\xspace}
\newcommand{\gev}{\ensuremath{{\mathrm{\,Ge\kern -0.1em V}}}\xspace}
\newcommand{\tev}{\ensuremath{{\mathrm{\,Te\kern -0.1em V}}}\xspace}

\newcommand{\m}[1]{\ensuremath{\mathcal{M}_{#1}}}
\newcommand{\mi}[1]{\ensuremath{\left(\mathcal{M}^{-1}\right)_{#1}}}
\newcommand{\B}[1]{\ensuremath{\mathcal{B}_{#1}}}
\newcommand{\Bp}[1]{\ensuremath{\mathcal{B}'_{#1}}}
\newcommand{\PP}[1]{\ensuremath{\mathcal{P}_{#1}}}
\newcommand{\Pp}[1]{\ensuremath{\mathcal{P}'_{#1}}}
\newcommand{\R}[1]{\ensuremath{\text{Reg}_{#1}}}
\newcommand{\Rp}[1]{\ensuremath{\text{Reg}'_{#1}}}

\begin{document}

\title{What is the right formalism to search for resonances?\\
II. The pentaquark chain}

\author{A.~Pilloni}
\email{pillaus@jlab.org}
\affiliation{Theory Center, Thomas Jefferson National Accelerator Facility,
Newport News, VA 23606, USA}

\author{J.~Nys}
\email{jannes.nys@ugent.be}
\affiliation{Theory Center, Thomas Jefferson National Accelerator Facility,
Newport News, VA 23606, USA}
\affiliation{Department of Physics and Astronomy, Ghent University, Belgium}
\affiliation{Center for Exploration of Energy and Matter, Indiana University, Bloomington, IN 47403, USA}
\affiliation{Physics Department, Indiana University, Bloomington, IN 47405, USA}

\author{M.~Mikhasenko}
\email{mikhail.mikhasenko@hiskp.uni-bonn.de}
\affiliation{{Universit\"at Bonn,
Helmholtz-Institut f\"ur Strahlen- und Kernphysik, 53115 Bonn, Germany}}

\author{M.~Albaladejo}
\affiliation{Departamento de F\'isica, Universidad de Murcia, E-30071 Murcia, Spain}

\author{C.~Fern\'andez-Ram\'irez}
\affiliation{Instituto de Ciencias Nucleares,
Universidad Nacional Aut\'onoma de M\'exico, Ciudad de M\'exico 04510, Mexico}

 \author{A.~Jackura}
 \affiliation{Center for Exploration of Energy and Matter,
 Indiana University, Bloomington, IN 47403, USA}
 \affiliation{Physics Department, Indiana University, Bloomington, IN 47405, USA}

 \author{V.~Mathieu}
\affiliation{Theory Center, Thomas Jefferson National Accelerator Facility, Newport News, VA 23606, USA}

 \author{N.~Sherrill}
 \affiliation{Center for Exploration of Energy and Matter,
 Indiana University, Bloomington, IN 47403, USA}
 \affiliation{Physics Department, Indiana University, Bloomington, IN 47405, USA}

\author{T.~Skwarnicki}
\affiliation{Syracuse University, Syracuse, NY 13244, USA}

\author{A.~P.~Szczepaniak}
\affiliation{Theory Center, Thomas Jefferson National Accelerator Facility,
Newport News, VA 23606, USA}
\affiliation{Center for Exploration of Energy and Matter,
Indiana University, Bloomington, IN 47403, USA}
\affiliation{Physics Department, Indiana University, Bloomington, IN 47405, USA}

\collaboration{Joint Physics Analysis Center}

\preprint{JLAB-THY-18-2700}


\begin{abstract}
We discuss the differences between several 
partial-wave analysis formalisms used 
in the construction of three-body decay amplitudes 
involving fermions. Specifically, we consider the decay
$\Lambdab \to \psi\,p \Km$, 
where the hidden charm pentaquark signal has been reported. 
We analyze the analytical properties 
of the amplitudes and separate kinematical 
and dynamical singularities. 
The result is an amplitude with the minimal 
energy dependence compatible with the $S$-matrix principles.
\end{abstract}

\maketitle

\section{Introduction}
\label{sec:intro}
In the recent years experiments such as \babar, Belle, BESIII,
CLAS, COMPASS, GlueX, LHCb, have produced significant amount of high-precision data on three-body hadron decays,
garnering information on new hadronic states
\cite{Esposito:2016noz,Lebed:2016hpi,Olsen:2017bmm,Guo:2017jvc,Karliner:2017qhf}.
 To put existence of such states on firm theoretical footing and to determine their physical properties rigorous amplitude analysis is needed. There are well established methods based on first principles of reaction theory for construction  
of reaction amplitudes describing three particle decays of hadrons~\cite{Jacob:1959at,Chung:1971ri,Collins:1977jy,cookbook,Chung:1993da,Chung:2007nn,Filippini:1995yc,Anisovich:2006bc,Anisovich:2011fc,Adolph:2015tqa}. It appears, however that there is  significant confusion as to the role of various approximations that these methods entail. 
In an earlier work~\cite{Mikhasenko:2017rkh},
we 
pointed out that, contrary to the common wisdom, 
differences among the various
approaches are dynamical rather than kinematical in nature, and we showed that the lore for the \ls formalism to be  nonrelativistic is unjustified.
As an example, we discussed the decay $B\to \psi \pi K$,
which shows 
nontrivial structures appearing in the
Belle and LHCb data in 
$\psi(2S) \,\pi$~\cite{Mizuk:2009da,Chilikin:2013tch,Aaij:2015wza,Aaij:2015zxa},
and 
$J/\psi \,\pi$ channels~\cite{Chilikin:2014bkk}.
In the present paper, we 
extend the discussion to the more complicated fermion-boson case.
Our main goal is to properly separate  kinematical from dynamical  singularities.
In general, the analysis of kinematical singularities of amplitudes with fermions has to be handled with particular care, because of the additional branch point at vanishing value of the Mandelstam variables~\cite{Cohen-Tannoudji:1968kvr}, and because fermions and antifermions have opposite intrinsic parities. Hence, one expects different behavior of the amplitudes at threshold and pseudothreshold.
We thus believe that study of such amplitudes deserves an extended discussion. Moreover, because of the possible existence of hidden charm pentaquarks, there is particular interest in final states containing the nucleon, a light meson and a charmonium~\cite{Lebed:2016hpi,Esposito:2016noz,Olsen:2017bmm}.
 In this paper we thus study the amplitudes for the 
 reaction 
$\Lambdab \to \jpsi p \Km$ 
in which a prominent pentaquark-like signal in the $\jpsi p$
invariant mass observed at LHCb~\cite{Aaij:2015tga,Aaij:2016phn}.

The paper is organized as follows. In Sec.~\ref{sec:s.channel} we discuss the canonical approach used to analyze the  $\Lambdab\to \psi p \Km$ decay.  By relating the helicity partial waves to the Lorentz scalar amplitudes via the partial-wave expansion, we derive constraints on the amplitudes and isolate the kinematical singularities. The results, and the comparison with the  \ls partial-wave amplitudes, are summarized in Sec.~\ref{sec:results}. In Sec.~\ref{sec:macdowell}, we focus on the mass dependence of our solution, and the singularities at $s=0$.
In Sec.~\ref{sec:comparison} we examine the Covariant Projection Method (\cpt) approach and compare it to our results.
Conclusions are given in Sec.~\ref{sec:sc}.
For ease of readability in the main text,
most of the technical details are given in the appendices where we also give a practical parameterization of the amplitudes suitable for data analysis.

\section{Analyticity constraints for \texorpdfstring{$\Lambdab \to \psi p \Km$}{Lambda\_b -> psi p K}}
\label{sec:s.channel}

\begin{figure}[t]
\centering
\subfigure[\ Decay]{
\includegraphics[]{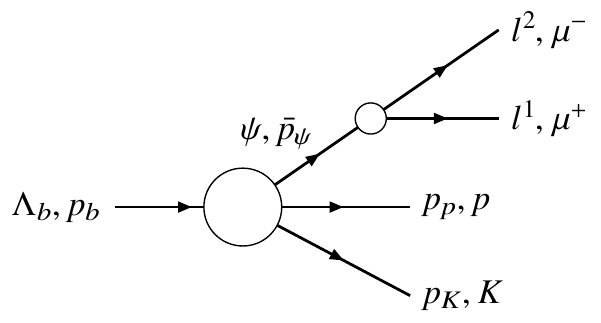} \label{fig:decay.diag1}}
\hspace{2cm}
\subfigure[\ $s$-channel scattering]{
\includegraphics[]{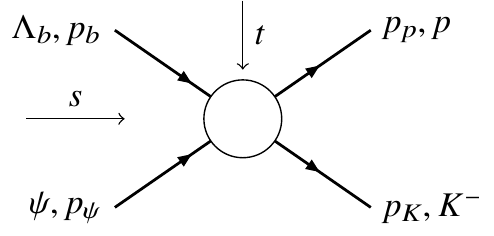}\label{fig:decay.diag2}}
\caption{
Reaction diagrams for (a) the $\Lambda_b\to \psi(\to\mu^-\mu^+) p \Km$ decay process, and for (b)
 the $\Lambda_b \psi \to p \Km$ $s$-channel scattering process. 
 } \label{fig:decay.diag}
\end{figure}

In Fig.~\ref{fig:decay.diag} we
specify the kinematics for the decay $\Lambdab \to \psi(\to \mu^+\mu^-) p \Km$. In the following, we follow the arguments presented in ~\cite{Mikhasenko:2017rkh}. 
We will be able to identify and characterize all kinematical singularities as either pertaining to (pseudo)thresholds, or to the vanishing of particles' energies.  The particles $\Lambdab$, $p$, and $\Km$ are stable against the strong interaction, and the $\psi$ is narrow enough, allowing one to factorize its decay dynamics. Thus, we focus on the amplitude in which $\psi$ is  also considered as stable.
In the following, we analyze the equivalent scattering problem $\Lambdab \psi \to p \Km$, and we refer to~\cite{Mikhasenko:2017rkh} for an extended discussion of  crossing symmetry.
We use $p_i$, $i=1$, $2$, $3$, and $4$ to label the momenta of $\Lambdab$, $\psi$, $p$, and $\Km$ respectively. We call $\pbarpsi = -\ppsi$ the momentum of the $\psi$ in the decay kinematics.
The helicity amplitude is denoted by $\helamp(s,t)$,
where
$\lambdapp$, $\lambdab$ and $\lambdapsi$ are the helicities of $p$, $\Lambdab$ and $\psi$, respectively.
The amplitude depends on the standard Mandelstam variables $s = (\ppp+\pK)^2$, $t = (\pb -\ppp)^2$, and $u = (\pb - \pK)^{2}$ with $s + t + u = \sum_i m_i^2$.

The $\Lambdab$ baryon decays weakly, so $\helamp$
is given by the sum of a parity conserving (PC) and a parity violating (PV) amplitudes.

We discuss here the PC amplitude in the $s$-channel, and we refer to Appendix~\ref{app:PV} for the summary of the PV amplitude. The $s$-channel resonances correspond to the $\Lambda^*$'s and dominate the reaction~\cite{Fernandez-Ramirez:2015tfa}. As discussed in the previous section, the analysis of the experimental data indicates a possible signal of resonances in the exotic $\psi p$ spectrum,  which in our notation correspond to the $u$-channel.

\begin{figure}[t]
  \centering
  \includegraphics[]{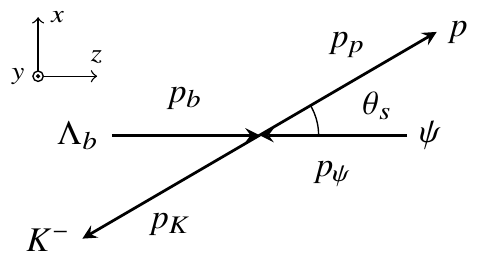}
  \caption{Scattering kinematics in the $s$-channel rest frame. In the decay kinematics, the momentum and the spin of the $\psi$ is reversed to keep the same helicity.}
  \label{fig:angles}
\end{figure}

In the center of mass of the $s$-channel scattering process, the momentum $\pb$ defines the $z$-axis, the momenta $\ppp$ and $\pK$ lie in the $xz$-plane, $p$ and $q$ denote magnitudes of relative momenta in the incoming ($\Lambda_b$, $\psi$) and the outgoing ($p$, $\Km$) states. The scattering angle $\theta_s$ is the polar angle of the proton (see Fig.~\ref{fig:angles}). The quantities are expressed
through the Mandelstam invariants,
\begin{equation} 
  \label{eq:zs}
 z_s \equiv \cos \theta_s =\frac{s(t-u)+(\mb^2-\mpsi^2)(\mpp^2-\mK^2)}{4s \,p q}
 \equiv\frac{n(s,t)}{p q},\qquad
 p = \frac{\lambdai^{1/2}}{2\sqrt{s}}, \qquad
 q = \frac{\lambdaf^{1/2}}{2\sqrt{s}},
\end{equation}
with $\lambda_{ik} \equiv \left(s -  (m_i+m_k)^2\right)\left(s -  (m_i-m_k)^2\right)$.
The function $4s\, n(s,t)$ is a polynomial in $s$ and $t$.\footnote{Note that the definition of $n(s,t)$ given here differs from the one used in~\cite{Mikhasenko:2017rkh} by the factor $4s$.}
To incorporate resonances in the $p \Km$ system with a certain spin $j$, we expand the amplitude in partial waves,
\begin{equation}
  \label{eq:helicity-pws.s-channel}
   \helamp = \frac{1}{4\pi}\sum_{j=M}^\infty (2j+1) \helamp^{j}(s) \,d_{\lambda,\lambda^\prime}^j(z_s),
\end{equation}
where $\helamp^j(s)$ 
are the helicity partial-wave amplitudes in the $s$-channel, 
$\lambda = \lambdab - \lambdapsi$, 
$\lambdap = \lambdapp$, and 
$M= \max(|\lambda|,|\lambdap|)$~\cite{Collins:1977jy}. 
We use the definition of the Wigner $d$ function as in ~\cite{Martin:1970}, \ie $d_{\lambda\lambda^\prime}^j(\cos\theta) = \left\langle j, \lambda | \exp\left(-i J_y \theta\right) | j,\lambda^\prime\right\rangle$,
that differs from the one in~\cite{Collins:1977jy} 
by $\theta \to  - \theta$. This results in a difference in our definition of the parity conserving helicity amplitudes given in  Eq.~\eqref{eq:PCHA} below.

Instead of working with an infinite number of helicity partial waves, 
we will consider the isobar model, customarily used in data analysis.\footnote{We remark that our discussion would be unchanged if applied to the untruncated partial-wave series.} The dynamical singularities in $s$, $t$ and $u$ are taken into account explicitly by a sum of different terms,
 \begin{equation}
  \label{eq:isobar.model}
  \helamp(s,t,u) = \heliso^{(s)}(s,t,u) + \heliso^{(t)}(s,t,u) + \heliso^{(u)}(s,t,u),
\end{equation}
with
\begin{equation}
\label{eq:isobar}
\heliso^{(s)}(s,t,u) = \frac{1}{4\pi}\sum_{j=M}^{J_\text{max}} (2j+1) \heliso^{(s) j}(s) \,d_{\lambda,\lambdap}^j(z_s),
\end{equation}
and $J_\text{max} < \infty$. In this model, it is assumed that the entire dynamical information is expressed by the isobar amplitudes, which are functions of a single Mandelstam variable: $A^{(x)j} = A^{(x)j}(x)$, with $x=s,t,u$.
The expressions for the $(t)$ and $(u)$ isobars are similar to Eq.~\eqref{eq:isobar}. In the following we focus on the $s$-channel isobars, and drop the $(s)$ superscript. The $u$-channel isobars are described in the appendices~\ref{app:uchannel},~\ref{app:uchannelpv}).

In Eqs.~\eqref{eq:helicity-pws.s-channel},\eqref{eq:isobar} the entire $t$ dependence originates from the $d$ functions. Specifically,  the $d$ functions have singularities in $z_s$ which lead to kinematical singularities in $t$. We define the functions
\begin{equation}
\hat d_{\lambda\lambdap}^j(z_s) = \frac{d_{\lambda\lambdap}^j(z_s)}{\xi_{\lambda\lambdap}(z_s)} ,
\end{equation}
with
\begin{equation}\label{eq:half_angle_factor}
\qquad\xi_{\lambda\lambdap}(z_s) = \left(\sqrt{1-z_s}\right)^{|\lambda - \lambdap|}\left(\sqrt{1+z_s}\right)^{|\lambda+\lambdap|}=\left(\sqrt2 \sin\tfrac{\theta_s}{2}\right)^{|\lambda - \lambdap|}\left(\sqrt 2\cos\tfrac{\theta_s}{2}\right)^{|\lambda+\lambdap|},
\end{equation}
being the so-called half angle factor that contains  all the kinematic singularities in $t$. The reduced rotational function $\hat d_{\lambda\lambdap}^j(z_s)$ is a polynomial of $n(s,t)/pq$
of order $j - M$, see Eq.~\eqref{eq:zs}.
For $\lambda,\lambdap \ne0$, the functions $d^j_{\lambda\lambdap}(z_s)$ have no definite parity. 
This means that the product $(pq)^{j-M} \hat d^j_{\lambda\lambdap}(z_s)$ contains terms with odd powers of $pq$ that still have  kinematic branch-point singularities in $s$. To be able to remove these singularities from the amplitude, we need to define the so-called parity-conserving helicity amplitudes (PCHAs),
\begin{align}
\label{eq:PCHA}
\heliso^\eta(s,t) = \frac{\heliso(s,t)}{ \xi_{\lambda\lambda^\prime}(z_s)} + \eta \, \eta_{\psi}\, \eta_b (-1)^{\lambda^\prime - M} \frac{A_{\lambdapp, -\lambdab -\lambdapsi}(s,t)}{\xi_{-\lambda\lambda^\prime}(z_s)},
\end{align}
where $\eta_\psi = \eta_b = +$ are the naturality of the $\psi$ and $\Lambdab$, respectively. These functions are free of kinematical singularities in $t$. Similarly, we can split the isobars into natural and unnatural ones 
\begin{align}
\heliso^j(s) = \heliso^{j+}(s) + \heliso^{j-}(s).
\end{align}
where we defined the definite-parity partial-wave/isobar  amplitudes,
\begin{align}
A_{\lambdapp,-\lambdab -\lambdapsi}^{j\eta }(s) = \eta \, \eta_\psi \, \eta_b \, \heliso^{j\eta }(s) \label{eq:definite_parity_partial_waves}.
\end{align}
We introduce the definite-parity
Wigner $d$ functions by,
\begin{equation}
\hat d^{j\eta}_{\lambda\lambdap}(z_s) = \hat d^{j}_{\lambda\lambdap}(z_s) + \eta(-1)^{\lambdap - M}\hat d^{j}_{-\lambda\lambdap}(z_s).
\end{equation} 
One can check that the function $\hat d^{j+}_{\lambda\lambdap}(z_s)$ is a definite-parity polynomial of order $j-M$, \ie $\hat d^{j+}_{\lambda\lambdap}(-z_s) = (-1)^{j-M} \hat d^{j+}_{\lambda\lambdap}(z_s)$. \footnote{Note that at leading order in $z_s$, $\xi_{\lambda \lambdap} (z_s) \propto z_s^M$.} Similarly, $\hat d^{j-}_{\lambda\lambdap}(z_s)$ is a definite parity polynomial of order $j-M-1$, and therefore subleading in the $z_s \to \infty$ limit. We refer the reader to Appendix~\ref{app:wigner} for a more detailed discussion on the (sub)leading behavior of the $\hat d^{j(\eta)}_{\lambda\lambdap}(z_s)$.
In terms of these isobars, the PCHAs read
\begin{align}
\heliso^\eta(s,t) &= \frac{1}{4\pi} \sum_{j=M} (2j+1) \Bigg[ \left(\heliso^{j\eta}(s) + \heliso^{j-\eta}(s)\right) \hat  d^j_{\lambda \lambda^\prime}(z_s)\nonumber \\&\qquad\qquad\qquad+ \eta \, \eta_{\psi} \, \eta_b (-1)^{\lambda^\prime - M} \left(A_{\lambdapp,-\lambdab-\lambdapsi}^{j\eta}(s) + A_{\lambdapp,-\lambdab-\lambdapsi}^{j-\eta}(s)\right) \hat d^j_{-\lambda \lambda^\prime}(z_s)\Bigg]\nonumber\\
&= \frac{1}{4\pi} \sum_{j=M} (2j+1) \Bigg[ \left(\heliso^{j\eta}(s) + \heliso^{j-\eta}(s)\right) \hat d^j_{\lambda \lambda^\prime}(z_s) \nonumber\\&\qquad\qquad\qquad
+ (-1)^{\lambda^\prime - M} \left(A_{\lambdapp,\lambdab\lambdapsi}^{j\eta}(s) - A_{\lambdapp,\lambdab\lambdapsi}^{j-\eta}(s)\right) \hat d^j_{-\lambda \lambda^\prime}(z_s)\Bigg]\nonumber\\
&= \frac{1}{4\pi} \sum_{j=M} (2j+1) \Bigg[ \heliso^{j\eta}(s)  \hat d^{j+}_{\lambda \lambda^\prime}(z_s)
+ A_{\lambdapp,\lambdab\lambdapsi}^{j-\eta}(s) \hat d^{j-}_{\lambda \lambda^\prime}(z_s)\Bigg],\label{eq:definite_parity_decomposition}
\end{align}
where we applied the parity relations in Eq.~\eqref{eq:definite_parity_partial_waves} in the transition from the first to second line.
We note that, for given $\eta$,  isobars with both naturalities contribute to the $\heliso^\eta(s,t)$.\footnote{The name ``Parity-Conserving Helicity Amplitudes'' arises from Regge theory, where in the limit $t \to \infty$ (which implies $z_s \to \infty$) the contribution from the opposite naturality, $-\eta$, is proportional to $\hat d^{j-}_{\lambda\lambdap}(z_s)$ which is negligible compared to $\hat d^{j+}_{\lambda\lambdap}(z_s)$. In the case at hand, however, we consider the limits $q \to 0$ or $p \to 0$, where the kinematic factors of the partial-wave amplitudes are also relevant to determine the leading behavior of the two contributions in Eq.~\eqref{eq:definite_parity_decomposition}.}
The helicity isobars  $\heliso^j(s)$ have singularities in $s$, which have both  dynamical and kinematical origin.
The kinematical singularities in $s$, just like the $t$-dependent kinematical singularities, arise because of particle spin.
We explicitly isolate the kinematic factors in $s$,
and denote the kinematical singularity-free helicity isobar  amplitudes by $\hheliso^j(s)$.
First, we take out the factor $(p q)^{j - M}$ from the $\heliso^j(s)$.
This factor cancels the threshold and pseudothreshold singularities in $s$ that appear in $\hat d^j_{\lambda\lambdap}(z_s)$.
Second, we follow~\cite{Collins:1977jy} and introduce the additional kinematic factor $K^\eta_{MN}$. These factors are required to account for the mismatch between the $j$ and $L$ dependence in the angular momentum barrier factors in
the
presence of particles with spin. Specifically, it is expected that $\heliso^{j\eta}(s) \sim p^{L_1}$ ($\heliso^{j\eta}(s) \sim q^{L_2}$) at $\Lambda_b \psi$-threshold ($p \Km$-threshold), where $L_1$ and $L_2$ are the lowest possible orbital angular momenta in the given helicity and parity combination.
The definite-parity, kinematical-singularity-free helicity isobar amplitudes $\hheliso^{j \eta}(s)$ are defined by 
\begin{subequations}
\begin{align}
A^{j\eta}_{\lambdapp, \lambdab \lambdapsi} & \equiv K^\eta_{MN} (pq)^{j-M}\hat A^{j\eta}_{\lambdapp, \lambdab \lambdapsi} & &\text{for }j\ge \tfrac{3}{2} ,\label{eq:Ahatdefschannel}\\
A^{1/2,\eta}_{\lambdapp, \lambdab \lambdapsi} & \equiv  \left(\frac{p\sqrt{s}}{\mpsi}\right)^{1+\eta} \left(\mmf{1/2}\right)^{-\eta} K^\eta_{1/2,1/2} \hat A^{1/2,\eta}_{\lambdapp, \lambdab \lambdapsi}
&&\text{for }j={\tfrac{1}{2}} \text{ and }M=\tfrac{1}{2}\label{eq:Kfactorshalfnat},\\
A^{1/2,\eta}_{\lambdapp, \lambdab \lambdapsi} & \equiv 0 &&\text{for }j={\tfrac{1}{2}} \text{ and }M=\tfrac{3}{2},
\end{align}
\end{subequations}
with $N= \min(|\lambda|,|\lambdap|) = \tfrac{1}{2}$, and
\begin{subequations}\label{eq:KpmMN}
\begin{align}
 K^+_{MN} &=\left(\frac{p\sqrt{s}}{\mpsi}\right)^{M - \frac{3}{2}}
\left(\frac{q\sqrt{s}}{\mpp}\right)^{M + \frac{1}{2}} \left(\frac{1}{\sqrt{s}}\right)^{M - N} \mmf{+}, \\
 K^-_{MN} &=
 \left(\frac{p\sqrt{s}}{\mpsi}\right)^{M - \frac{1}{2}}
\left(\frac{q\sqrt{s}}{\mpp}\right)^{M - \frac{1}{2}} \left(\frac{1}{-\sqrt{s}}\right)^{M - N} \mmf{-},
\end{align}
\end{subequations}
where the $\mmf{\pm,1/2}$ are regular functions for $\sqrt{s}>0$. The functional form of the latter will be discussed in detail in Section~\ref{sec:macdowell}. In addition, the $K$-factors have powers of $\sqrt{s}$ as required to ensure factorization of the isobar amplitude into contributions from
distinct vertices~\cite{Collins:1977jy}. 

The isobar amplitudes $\hheliso^{j,\eta}(s)$ contain the dynamical information of the model. Often they are parameterized in terms of a sum of Breit-Wigner amplitudes with Blatt-Weisskopf barrier factors.

Once we have removed the kinematic singularities from the isobar amplitudes and the corresponding angular functions, we are now in a position to remove the singularities from the full amplitude. Therefore, we take out the factor $K_{MN}^\eta$ and define the amplitudes $F$ which are kinematic singularity-free PCHAs (KSF-PCHAs),
\begin{align}
F_{\lambdapp,\lambdab\lambdapsi}^\eta(s,t) & \equiv  \frac{1}{K^\eta_{MN}}\heliso^\eta(s,t) \nonumber \\
&= \frac{1}{4\pi} \sum_{j=3/2} (2j+1)(pq)^{j-M} \Bigg[ \hheliso^{j\eta}(s)  \hat d^{j+}_{\lambda \lambda^\prime}(z_s)
+ \hheliso^{j-\eta}(s) \frac{K^{-\eta}_{MN}}{K^\eta_{MN}}\hat d^{j-}_{\lambda \lambda^\prime}(z_s)\Bigg]\nonumber\\
&\quad + \frac{1}{2\pi} \hheliso^{1/2,\eta}(s)  \left(\frac{p\sqrt{s}}{\mpsi}\right)^{1 + \eta} \left(\mmf{1/2}\right)^{-\eta} \sqrt{2}\, (-1)^{\tfrac{1}{2} \left( |\lambda - \lambdap| + \lambda - \lambdap\right)} \delta_{\abs{\lambda},1/2},
\label{eq:ksf-pcha}
\end{align}
where the ratio $K^{-\eta}_{MN}/K^\eta_{MN} = (-)^{M-N}\left(p \mpp \big/ q \mpsi\right)^{\eta} \mmf{-\eta}/\mmf{\eta}$. While the KSF-PCHAs are free of kinematical singularities, they are not necessarily independent for all kinematics. Indeed, we will illustrate below that
additional constraints must be fulfilled by the isobar amplitudes for certain kinematics.
Therefore, as in~\cite{Mikhasenko:2017rkh}, we seek a representation of $\heliso(s,t)$ in terms of a set of covariant structures that explicitly account for the kinematic part of the amplitude.
For the PC amplitude, the basis with minimal energy dependence is given by
\begin{equation}
\heliso(s,t)  = \epsilon_\mu(\ppsi,\lambdapsi) \, \bar{u}(\ppp,\lambdapp) \left(\sum_{i=1}^6 C_i(s,t)\, M_i^\mu \right) u(\pb,\lambdab), \label{eq:cov}
\end{equation}
with
\begin{subequations}
\label{eq:cgln}
\begin{align}
M_1^\mu &= \gamma^5 \, \pb^\mu, &
M_2^\mu &= \gamma^5 \, \ppp^\mu, &
M_3^\mu &= \gamma^5 \,\slashedppsi\, \pb^\mu,\\
M_4^\mu &= \gamma^5 \,\slashedppsi\, \ppp^\mu, &
M_5^\mu &= \gamma^5 \gamma^\mu, &
M_6^\mu &= \gamma^5 \,\slashedppsi\, \gamma^\mu.
\end{align}
\end{subequations}
 and the scalar functions $C_i(s,t)$ are free from kinematical singularities.
There are six independent $M_{1\dots6}^{\mu}$ tensors, and any other possible combination can be reduced to these using the Dirac equation for the spinors, or the orthogonality relation $\epsilon_\mu(\ppsi,\lambdapsi)\ppsi^\mu = 0$.
Alternatively, one can use the CGLN basis defined in~\cite{Berends:1967vi}, for pseudoscalar-meson electro-production. However, these covariant structures enforce a gauge-invariance principle which does not apply here since $\psi$ is a massive vector particle. 
 Had we used the CGLN basis, there would be unnecessary kinematic zeros. The PC amplitude requires a $\gamma^5$ because of the unnatural $\Km$ parity.
The explicit expressions for the polarization vectors and spinors are given in Appendix~\ref{app:polarization}.
We can match Eq.~\eqref{eq:ksf-pcha} and~\eqref{eq:cov}, and express the scalar functions as a sum over kinematical singularity free helicity isobars. This yields
\begin{equation}
\label{eq:prima}
\begin{pmatrix}  F^+_{+,++} \\ F^+_{+,+0} \\ F^+_{+,+-} \\ F^-_{+,++} \\ F^-_{+,+0} \\ F^-_{+,+-}\end{pmatrix}
=  \frac{\sqrt{\Eb + \mb}}{\sqrt{\Epp + \mpp}}\frac{1}{\mmf{+}} \mathcal{M}
 \begin{pmatrix} C_1 \\ C_2 \\ C_3 \\ C_4 \\ C_5 \\ C_6\end{pmatrix},
\end{equation}
with $\mathcal{M}$ a $6\times 6$ matrix that encodes all the kinematic factors and is provided in Appendix~\ref{app:matrices}.\footnote{$F^+_{+,++}$ stands for $F^+_{\lambdapp=+\tfrac{1}{2},\lambdab=+\tfrac{1}{2}, \lambdapsi = +1}$, and so on.} The factors $\sqrt{\Epp + \mpp}$ and $\sqrt{\Eb + \mb}$ are factored out to simplify the expression for $\mathcal{M}$. We stress that they have only singularities at $s=0$. For example,
\begin{equation} 
\label{eq:eplusm}
\sqrt{\Eb + \mb} = \sqrt{\frac{\left(\sqrt{s} + \mb - \mpsi\right)\left(\sqrt{s} + \mb + \mpsi\right)}{2\sqrt{s}}},
\end{equation}
and the physical region of $\sqrt{s}$ corresponds to  $\re \sqrt{s} > 0$. For $\mb > \mpsi$, which is the case here, the first factor in Eq.~\eqref{eq:eplusm} is always positive, and the only singularity is due to the branch point at $s=0$. This would be different if the fermion was lighter than the boson. In that case, the factor will have a singularity at pseudothreshold that has to be considered separately. The relation in Eq.~\eqref{eq:prima} can be inverted, leading to
\begin{equation}
\label{eq:grossa}
 \begin{pmatrix} C_1 \\ C_2 \\ C_3 \\ C_4 \\ C_5 \\ C_6\end{pmatrix}
= \frac{\sqrt{\Epp + \mpp}}{\sqrt{\Eb + \mb}} \mmf{+} \mathcal{M}^{-1}  \begin{pmatrix}  F^+_{+,++} \\ F^+_{+,+0} \\ F^+_{+,+-} \\ F^-_{+,++} \\ F^-_{+,+0} \\ F^-_{+,+-}\end{pmatrix}= \frac{\sqrt{\Epp + \mpp}}{\sqrt{\Eb + \mb}}\mmf{+} \left(\frac{1}{p^2} \mathcal{B} + \text{Reg}  \right)\begin{pmatrix}  F^+_{+,++} \\ F^+_{+,+0} \\ F^+_{+,+-} \\ F^-_{+,++} \\ F^-_{+,+0} \\ F^-_{+,+-}\end{pmatrix},
\end{equation}
where the matrices $\mathcal{B}$ and $\text{Reg}$ are regular at $p=0$. The explicit expression for the $\mathcal{M}^{-1}$ and the $\mathcal{B}$ matrices are in Appendix~\ref{app:matrices}. We
just report a few terms here to ease the discussion,
\begin{equation}
\label{eq:matricione}
\mathcal{B} =
\begin{pmatrix}
\B{11} & \B{12} & \B{13} & \B{14} & \B{15} & \B{16}\\
0 &0 & 0 & 0 & 0 & 0\\
\B{31} & \B{32} & \B{33} & \B{34} & \B{35} & \B{36}\\
0 & 0 & 0 & 0 & 0 & 0\\
\frac{\mpsi \left(\sqrt{s}-\mb\right)\left(\Eb + \mb\right)}{   4   \mpp \sqrt{s}} & 0 & \frac{ n(s,t) \left( \sqrt{s}-\mb \right)\left(\Eb + \mb\right)}{4  \mpp^2} & 0 & 0 & 0\\
\frac{\mpsi\left(\Eb + \mb\right)}{   4 \mpp   \sqrt{s}} & 0 & \frac{  n(s,t) \left(\Eb + \mb\right)}{    4  \mpp^2} & 0 & 0 & 0
\end{pmatrix}.
\end{equation}
Since the $C_i$ functions must be regular at $p=0$, \ie for $s = (\mb \pm \mpsi)^2 \equiv s_\pm$, the combinations of KSF-PCHAs $F_{\lambdapp,\lambdab\lambdapsi}^\eta(s,t)$  in Eq.~\eqref{eq:grossa} must conspire to cancel the $1/p^2$ pole.
This translates into a relation between the various isobar amplitudes  $\hheliso^{j\eta}(s)$. 
As an example, let us consider
the last two rows in Eq.~\eqref{eq:grossa}. Inspecting the matrix elements in Eq.~\eqref{eq:matricione} one 
finds that two emerging conditions
are not independent and lead to,
\begin{equation}
F^+_{+,++} + \frac{\sqrt{s}}{m_p m_\psi} n(s,t) F^+_{+,+-} \simeqzero{p^2},
\end{equation}
where we mean here is that this combination must vanish as $p^2$ for $p \to 0$. The conspiracy relation can be written in terms of the isobar amplitudes by inserting the expression for the $F$'s in terms of the isobars, given in Eq.~\eqref{eq:ksf-pcha}. Since the isobars of different spin are independent, we can consider each $j$ individually. For $j \ge \tfrac{3}{2}$ we obtain
\begin{align}
\label{eq:conspiracy1.natbefore}
&(pq)^{j-1/2}\Bigg[ \hat A_{+,++}^{j+}(s)\,  \hat d^{j+}_{-1/2,1/2}(z_s)
+ \hat A_{+,++}^{j-}(s) \frac{p\, \mpp }{q\, \mpsi}\frac{\mmf{-}}{\mmf{+}}\, \hat d^{j-}_{-1/2,1/2}(z_s)\Bigg]\nonumber \\
&\qquad+ (pq)^{j-1/2} \frac{\sqrt{s}}{m_p m_\psi}  \frac{n(s,t) }{pq}
\Bigg[ \hat A_{+,+-}^{j+}(s)\,  \hat d^{j+}_{3/2,1/2}(z_s)
- \hat A_{+,+-}^{j-}(s) \frac{p\, \mpp }{q\, \mpsi} \frac{\mmf{-}}{\mmf{+}}\, \hat d^{j-}_{3/2,1/2}(z_s)\Bigg]
\simeqzero{p^2}.
\end{align}
When $p \to 0$, $z_s \to \infty$ and the leading $\hat d^{j+}_{-1/2,1/2}(z_s)$ and $\hat d^{j+}_{3/2,1/2}(z_s)$ diverge as $1/p^{j-1/2}$ and $1/p^{j-3/2}$, respectively. This divergence is canceled by the threshold factor $(pq)^{j-1/2}$, but an additional relation between $\hat A_{+,++}^{j+}(s)$ and $\hat A_{+,+-}^{j+}(s)$ is needed to cancel the additional $1/p^2$ pole appearing in Eq.~\eqref{eq:grossa}. On the other hand, the subleading $\hat d^{j-}_{-1/2,1/2}(z_s)$ and $\hat d^{j-}_{3/2,1/2}(z_s)$ diverge as $1/p^{j-3/2}$ and $1/p^{j-5/2}$ only, and together with the additional factor of $p$ coming from the mismatch factors $K_{MN}^\eta$ and the threshold factor, vanish as $p^2$ to cancel the $1/p^2$ pole. Therefore, the opposite-naturality waves do not contribute to this type of  conspiracy relations. It is also straightforward to check that the expressions are regular when $q\to 0$. One can use the asymptotic expansion of the Wigner $d$ functions (the full expressions are in Appendix~\ref{app:wigner}),
\begin{align}
\hat d^{j+}_{-1/2,1/2}(z_s) &\sim \frac{z_s^{j-1/2}f(j)}{\left\langle \frac{1}{2},\frac{1}{2};1,-1|\frac{3}{2},-\frac{1}{2}\right\rangle\left\langle \frac{3}{2},-\frac{1}{2};j-\frac{3}{2},0|j,-\frac{1}{2}\right\rangle}, &
 \hat d^{j+}_{3/2,1/2}(z_s) &\sim \frac{-z_s^{j-3/2} f(j)}{\left\langle \frac{3}{2},\frac{3}{2};j-\frac{3}{2},0|j,\frac{3}{2}\right\rangle}%
\end{align}
and reduce Eq.~\eqref{eq:conspiracy1.natbefore} to
\begin{align}
\label{eq:conspiracy1.nat}
\frac{ \hat A_{+,++}^{j+}(s)  }{\left\langle \frac{1}{2},\frac{1}{2};1,-1|\frac{3}{2},-\frac{1}{2}\right\rangle\left\langle \frac{3}{2},-\frac{1}{2};j-\frac{3}{2},0|j,-\frac{1}{2}\right\rangle}
- \frac{\sqrt{s}}{\mpp \mpsi} \frac{ \hat A_{+,+-}^{j+}(s) }{\left\langle \frac{3}{2},\frac{3}{2};j-\frac{3}{2},0|j,\frac{3}{2}\right\rangle} =0.
\end{align}
We now examine the conditions that emerge for the first and third rows in Eq.~\eqref{eq:matricione}. These  involve both natural and unnatural isobars. Although strictly speaking  conspiracy relations might be realized by complicated cancellations involving all possible isobars, we again assume that isobars carrying different quantum numbers are independent. It is then possible to (i) break each one of the equations in Eq.~\eqref{eq:matricione} into separate equations for natural and unnatural isobars, and (ii)  break them further by counting the powers of $z_s$. This leads to the following conditions,
\begin{align}
\hat A^{j-}_{+,++}(s) \hat d^{j+}_{-1/2,1/2}(z_s) -
 \frac{\sqrt{2}  \left(\sqrt{s} - \mb\right)}{ \mpsi  } \hat A^{j-}_{+,+0}(s) \hat d^{j+}_{1/2,1/2}(z_s)
+ \frac{ s-\mb \left(2 \Epsi + \mb\right)}{    \mpp  \mpsi^3 } \sqrt{s} \hat A^{j-}_{+,+-}(s)  \,z_s \hat d^{j+}_{3/2,1/2}(z_s) &\simeqzero{p^2},\\
\frac{  }{ } \hat A^{j+}_{+,++}(s) \hat d^{j+}_{-1/2,1/2}(z_s)
 -\frac{ \sqrt{2}\left(\mb + \sqrt{s}\right)}{ \mpsi } \hat A^{j+}_{+,+0}(s) \hat d^{j+}_{1/2,1/2}(z_s)
- \frac{2 \Epsi \mb - \mb^2 + s}{     \mpp \mpsi^3 } \sqrt{s} \hat A^{j+}_{+,+-}(s)\,  z_s \hat d^{j+}_{3/2,1/2}(z_s)&\simeqzero{p^2},\\
\left(\sqrt{s}-\mb\right)\hat A^{j-}_{+,++}(s) \hat d^{j+}_{-1/2,1/2}(z_s)
 -\sqrt{2} \mpsi \hat A^{j-}_{+,+0}(s) \hat d^{j+}_{1/2,1/2}(z_s)
 - \frac{\Eb - \Epsi - \mb }{    \mpp    \mpsi }\sqrt{s} \hat A^{j-}_{+,+-}(s) \,z_s \hat d^{j+}_{3/2,1/2}(z_s) &\simeqzero{p^2},\\
 \left(\sqrt{s} + \mb\right) \hat A^{j+}_{+,++}(s) \hat d^{j+}_{-1/2,1/2}(z_s)
 - \sqrt{2} \mpsi \hat A^{j+}_{+,+0}(s) \hat d^{j+}_{1/2,1/2}(z_s)
+\frac{\Eb - \Epsi + \mb}{    \mpp \mpsi } \sqrt{s}\hat A^{j+}_{+,+-}(s) \, z_s \hat d^{j+}_{3/2,1/2}(z_s) &\simeqzero{p^2},
\end{align}
and using the asymptotic form of the $d$ functions, for the natural isobars, we obtain
\begin{align}\label{eq:conspiracy2.nat}
\frac{\hat A_{+,+0}^{j+}(s)}{\left\langle \frac{1}{2},\frac{1}{2};1,0|\frac{3}{2},\frac{1}{2}\right\rangle\left\langle \frac{3}{2},\frac{1}{2};j-\frac{3}{2},0|j,\frac{1}{2}\right\rangle}  = \frac{\Epsi}{\mpsi}\frac{\hat A_{+,++}^{j+}(s)}{\left\langle \frac{1}{2},\frac{1}{2};1,-1|\frac{3}{2},-\frac{1}{2}\right\rangle\left\langle \frac{3}{2},-\frac{1}{2};j-\frac{3}{2},0|j,-\frac{1}{2}\right\rangle}.
\end{align}

The conspiracy relation for the unnatural isobars are more cumbersome. We recall the relation between the helicity and the \ls couplings. To ease the notation, we will write only the initial state in the \ls form,\footnote{We remark that we used the convention $\left\langle S,\lambdab - \lambdapsi;L,0|j,\lambdab - \lambdapsi\right\rangle$ for the \ls Clebsch-Gordan coefficients. However, up to signs one can use $\left\langle L, 0; S,\lambdab - \lambdapsi|j,\lambdab - \lambdapsi\right\rangle$ to get equivalent results.}
\begin{equation}
 \label{eq:ls}
G_{\lambdapp,LS}^{j\eta}(s) = \sqrt{\frac{2L+1}{2j+1}} \sum_{\lambdab,\lambdapsi} {\textstyle\left\langle \frac{1}{2},\lambdab;1,-\lambdapsi|S,\lambdab-\lambdapsi\right\rangle\left\langle S,\lambdab - \lambdapsi;L,0|j,\lambdab - \lambdapsi\right\rangle}  A_{\lambdapp,\lambdab\lambdapsi}^{j\eta}(s).
\end{equation}
For the case at hand, this means
\begin{align}
A^{j-}_{+,+\lambdapsi} &= p^{j-1/2}q^{j-1/2}\Bigg[\sqrt{\frac{2j}{2j+1}} \Big(
\textstyle{\left\langle \frac{1}{2}, \frac{1}{2}; 1, -\lambdapsi \,\Big|\,\frac{1}{2}, \frac{1}{2} -\lambdapsi  \right\rangle\left\langle \frac{1}{2},\frac{1}{2}-\lambdapsi; j - \frac{1}{2},0\,\Big|\,j,\frac{1}{2}-\lambdapsi \right\rangle  } \, \hat G^{j-}_{j-1/2, 1/2}\nonumber\\
&\qquad+ \textstyle{ \left\langle \frac{1}{2}, \frac{1}{2}; 1, -\lambdapsi \,\Big|\,\frac{3}{2}, \frac{1}{2} -\lambdapsi  \right\rangle \left\langle \frac{3}{2},\frac{1}{2}-\lambdapsi; j - \frac{1}{2},0 \,\Big|\,j,\frac{1}{2}-\lambdapsi \right\rangle } \, \hat G^{j-}_{j-1/2, 3/2}\Big)\nonumber\\
&\qquad+ \sqrt{\frac{2j+4}{2j+1}}\textstyle{  \left\langle \frac{1}{2}, \frac{1}{2}; 1, -\lambdapsi \,\Big|\,\frac{3}{2}, \frac{1}{2} -\lambdapsi  \right\rangle\left\langle \frac{3}{2},\frac{1}{2}-\lambdapsi;j + \frac{3}{2},0 \,\Big|\,j,\frac{1}{2}-\lambdapsi \right\rangle} \, p^2 \hat G^{j-}_{j+3/2,3/2}\Bigg],
\end{align}
with $G^{j-}_{LS}(s) = p^L q^{j-1/2} \hat G^{j-}_{LS}(s)$. We remark that these relations hold for the $j=\tfrac{1}{2}$ case as well, and we do not need any separate consideration for it. There is only one \ls coupling with nonminimal $L$, which calls for one conspiracy equation only. However, the equations obtained from the first and third line in Eq.~\eqref{eq:matricione} give
\begin{subequations}
 \begin{align}
  F^-_{+,++} -
\frac{\sqrt{s} - \mb}{ \mpsi } \sqrt{2} \, F^-_{+,+0}
+ \frac{ n(s,t) \left(s-\mb \left(2 \Epsi + \mb\right)\right)}{    \mpp \mpsi^3 }\sqrt{s} F^-_{+,+-} &\simeqzero{p^2},\\
 \left( \sqrt{s}-\mb\right) \, F^-_{+,++}
 - \sqrt{2} \mpsi \,F^-_{+,+0}
  - \frac{ \left(\Eb - \Epsi - \mb\right) n(s,t) }{    \mpp    \mpsi }\sqrt{s} \,F^-_{+,+-}&\simeqzero{p^2},
\end{align}
\end{subequations}
and it is easy to check the two equations to be independent out of (pseudo)threshold. We evaluate the constraints at both threshold and pseudothreshold,

 \begin{align}
\frac{\hat A^{j-}_{+,++}}{\left\langle \frac{1}{2},\frac{1}{2};1,-1|\frac{1}{2},-\frac{1}{2}\right\rangle\left\langle \frac{1}{2},-\frac{1}{2};j-\frac{1}{2},0|j,-\frac{1}{2}\right\rangle} \mp
 \frac{\hat A^{j-}_{+,+0} }{\left\langle \frac{1}{2},\frac{1}{2};1,0|\frac{1}{2},\frac{1}{2}\right\rangle\left\langle \frac{1}{2},\frac{1}{2};j-\frac{1}{2},0|j,\frac{1}{2}\right\rangle}+ \frac{\sqrt{s_\pm }   }{    \mpp \mpsi }\frac{\hat A^{j-}_{+,+-} \,C}{\left\langle \frac{3}{2},\frac{3}{2};j-\frac{1}{2},0|j,\frac{3}{2}\right\rangle} &\simeqzero{p^2},
\end{align}
with
\begin{align}
C= \frac{\left\langle \frac{1}{2},\frac{1}{2};1,-1|\frac{3}{2},-\frac{1}{2}\right\rangle\left\langle \frac{3}{2},-\frac{1}{2};j-\frac{1}{2},0|j,-\frac{1}{2}\right\rangle}{\left\langle \frac{1}{2},\frac{1}{2};1,-1|\frac{1}{2},-\frac{1}{2}\right\rangle\left\langle \frac{1}{2},-\frac{1}{2};j-\frac{1}{2},0|j,-\frac{1}{2}\right\rangle}-\frac{\left\langle \frac{1}{2},\frac{1}{2};1,0|\frac{3}{2},\frac{1}{2}\right\rangle\left\langle \frac{3}{2},\frac{1}{2};j-\frac{1}{2},0|j,\frac{1}{2}\right\rangle}{\left\langle \frac{1}{2},\frac{1}{2};1,0|\frac{1}{2},\frac{1}{2}\right\rangle\left\langle \frac{1}{2},\frac{1}{2};j-\frac{1}{2},0|j,\frac{1}{2}\right\rangle}. \label{eq:C.clebsches}
\end{align}
By restoring the kinematic factors,
 \begin{align}\label{eq:conspiracy.unnat}
\frac{A^{j-}_{+,++}}{\left\langle \frac{1}{2},\frac{1}{2};1,-1|\frac{1}{2},-\frac{1}{2}\right\rangle\left\langle \frac{1}{2},-\frac{1}{2};j-\frac{1}{2},0|j,-\frac{1}{2}\right\rangle} \mp
 \frac{A^{j-}_{+,+0} }{\left\langle \frac{1}{2},\frac{1}{2};1,0|\frac{1}{2},\frac{1}{2}\right\rangle\left\langle \frac{1}{2},\frac{1}{2};j-\frac{1}{2},0|j,\frac{1}{2}\right\rangle}- \frac{ A^{j-}_{+,+-} \,C}{\left\langle \frac{3}{2},\frac{3}{2};j-\frac{1}{2},0|j,\frac{3}{2}\right\rangle} &\simeqzero{p^2}.
\end{align}
At threshold, this matches with the \ls constraint. To
interpolate with the pseudothreshold result, we replace \mbox{$\mp  \to -\Epsi/\mpsi$}.

To summarize, we used analyticity constraints to derive relations between the different helicity isobars. At threshold, these relations are in agreement with the expectations derived from the \ls decomposition. Similar constraints are derived at pseudothreshold. To interpolate between the two constraints, we add an energy dependent factor $\Epsi/\mpsi$ in the $\lambdapsi = 0$ amplitude. This results in the minimal kinematic dependence as required by analyticity.

\section{\texorpdfstring{The generic parameterization for the $s$-channel isobars}{The generic parameterization for the s-channel isobars}}
\label{sec:results}
In this section we derive a general parametrization for the isobar amplitude  which takes into account the conspiracy relations derive in the preceding section. 
A generic parameterization for the natural isobars which fullfills Eqs.~\eqref{eq:conspiracy1.nat} and~\eqref{eq:conspiracy2.nat} is given by
\begin{subequations}
\label{eq:general.form.nat}
 \begin{align}
\frac{\mpsi}{\mpp}\hat A_{+,++}^{j+}(s) &= {\textstyle\left\langle \frac{1}{2},\frac{1}{2};1,-1|\frac{3}{2},-\frac{1}{2}\right\rangle\left\langle \frac{3}{2},-\frac{1}{2};j-\frac{3}{2},0|j,-\frac{1}{2}\right\rangle} \, g_{j+}(s) + p^2 \,f_{j+}(s),\\
\frac{\mpsi}{\mpp}\hat A_{+,+0}^{j+}(s) &= {\textstyle\left\langle \frac{1}{2},\frac{1}{2};1,0|\frac{3}{2},\frac{1}{2}\right\rangle\left\langle \frac{3}{2},\frac{1}{2};j-\frac{3}{2},0|j,\frac{1}{2}\right\rangle} \, \frac{\Epsi}{\mpsi}   g'_{j+}(s) + p^2 \, f_{j+}'(s), \label{eq:AzeroEpsiovermpsi}\\
\frac{\mpsi}{\mpp}\hat A_{+,+-}^{j+}(s) &= {\textstyle\left\langle \frac{3}{2},\frac{3}{2};j-\frac{3}{2},0|j,\frac{3}{2}\right\rangle} \,\frac{\mpp \mpsi}{\sqrt{s}}\,  g''_{j+}(s) + \frac{\mpp \mpsi}{\sqrt{s}} p^2 \, f_{j+}''(s),\label{eq:genformnatc}
\end{align}
\end{subequations}
where $g^{(\prime,\prime\prime)}_{j+}(s)$ and $f^{(\prime,\prime\prime)}_{j+}(s)$ are regular functions at $s=s_\pm$, and $g_{j+}(s_\pm)=g'_{j+}(s_\pm)=g''_{j+}(s_\pm)$. 
 The branch points at $s=0$ are not constrained by the threshold relations. Their origin is dynamical and has to be addressed in the context of analyticity in $j$. We decided to include appropriate factors of $\sqrt{s}$ to make these formulae more similar to the \ls ones. For example, the additional factor of $\mpp\mpsi/\sqrt{s}$ in front of $f''_{j+}(s)$ in Eq.~\eqref{eq:genformnatc} is unconstrained by these relations, but it has been inserted by analogy  with \ls. Similarly, we decided the subleading $f_{j+}^{(\prime,\prime\prime)}(s)$ functions to appear with a factor $p^2$ insted of $\lambda_{b\psi}$, unlike in~\cite{Mikhasenko:2017rkh}. 

Upon restoration of the kinematic factors, the original helicity isobars amplitudes read ($j\ge \tfrac{3}{2}$)
\begin{subequations}
\label{eq:helicitypw.nat}
\begin{align}
A_{+,++}^{j+}(s) &= \mmf{+}p^{j-3/2} q^{j+1/2}\bigg[{\textstyle\left\langle \frac{1}{2},\frac{1}{2};1,-1|\frac{3}{2},-\frac{1}{2}\right\rangle\left\langle \frac{3}{2},-\frac{1}{2};j-\frac{3}{2},0|j,-\frac{1}{2}\right\rangle} \, g_{j+}(s) + p^2 \,f_{j+}(s)\bigg],\\
A_{+,+0}^{j+}(s) &= \mmf{+}p^{j-3/2} q^{j+1/2}\bigg[{\textstyle\left\langle \frac{1}{2},\frac{1}{2};1,0|\frac{3}{2},\frac{1}{2}\right\rangle\left\langle \frac{3}{2},\frac{1}{2};j-\frac{3}{2},0|j,\frac{1}{2}\right\rangle} \, \frac{\Epsi}{\mpsi}   g'_{j+}(s) + p^2 \, f_{j+}'(s)\bigg]~,\label{eq:helicitypw.nat.zero}\\
A_{+,+-}^{j+}(s) &= \mmf{+} p^{j-3/2} q^{j+1/2}\bigg[{\textstyle\left\langle \frac{3}{2},\frac{3}{2};j-\frac{3}{2},0|j,\frac{3}{2}\right\rangle}  g''_{j+}(s) + p^2 \, f_{j+}''(s)\bigg].
\end{align}
\end{subequations}
A particular choice of the functions $g^{(\prime,\prime\prime)}_{j+}(s)$ and $f^{(\prime,\prime\prime)}_{j+}(s)$ constitutes
a given hadronic model.
For $j = \tfrac{1}{2}$, no conspiracy constraint is needed. Indeed, the isobars $\hheliso^{j+}(s)$ always appears with an additional factor $\propto p^2$, as shown in Eq.~\eqref{eq:Kfactorshalfnat}, and the pole is automatically canceled.

We can immediately cast this expression in the \ls basis.
For the natural isobars considered, this reads
\begin{align}
A^{j+}_{+,+\lambdapsi}(s) &= p^{j-3/2}q^{j+1/2}\Bigg[\sqrt{\frac{2j-2}{2j+1}}\textstyle{   \left\langle \frac{1}{2}, \frac{1}{2}; 1, -\lambdapsi \,\Big|\,\frac{3}{2}, \frac{1}{2} -\lambdapsi  \right\rangle\left\langle \frac{3}{2},\frac{1}{2}-\lambdapsi ; j - \frac{3}{2},0\,\Big|\,j,\frac{1}{2}-\lambdapsi \right\rangle} \,\hat G^{j+}_{j-3/2, 3/2}(s) \nonumber \\
&\qquad+ \sqrt{\frac{2j+2}{2j+1}} p^2 \Big(\textstyle{  \left\langle \frac{1}{2}, \frac{1}{2}; 1, -\lambdapsi \,\Big|\,\frac{1}{2}, \frac{1}{2} -\lambdapsi  \right\rangle\left\langle \frac{1}{2},\frac{1}{2}-\lambdapsi;j + \frac{1}{2},0 \,\Big|\,j,\frac{1}{2}-\lambdapsi \right\rangle} \,\hat G^{j+}_{j+1/2, 1/2}(s) \nonumber \\
&\qquad\qquad\qquad\qquad+ \textstyle{ \left\langle \frac{1}{2}, \frac{1}{2}; 1, -\lambdapsi \,\Big|\,\frac{3}{2}, \frac{1}{2} -\lambdapsi  \right\rangle \left\langle \frac{3}{2},\frac{1}{2}-\lambdapsi ;  j + \frac{1}{2},0\,\Big|\,j,\frac{1}{2}-\lambdapsi \right\rangle }\, \hat G^{j+}_{j+1/2,3/2}(s)\Big)\Bigg],\label{eq:mishachepalle}
\end{align}
with $G^{j+}_{LS}(s) = p^L q^{j+1/2} \hat G^{j+}_{LS}(s)$, and the dependence on $\lambdapp$ is understood. We remark that there are two \ls couplings having nonminimal $L$ in Eq.~\eqref{eq:mishachepalle}, and at $p=0$ the three helicity couplings will depend on one independent \ls coupling only. This will require indeed two equations to be satisfied,  \ie Eqs.~\eqref{eq:conspiracy1.nat} and~\eqref{eq:conspiracy2.nat}. This expression matches Eq.~\eqref{eq:general.form.nat} by identifying
\begin{subequations}
\label{eq:lsmatchingnat}
\begin{align}
g_{j+}(s) &= \sqrt{\frac{2j-2}{2j+1}} \, \frac{1}{\mmf{+}}\hat{G}^{j+}_{j-3/2,3/2}(s), \\
f_{j+}(s) &= \sqrt{\frac{2j+2}{2j+1}} \frac{1}{\mmf{+}}\Big(\textstyle{\left\langle \frac{1}{2}, \frac{1}{2}; 1, -1 \,\Big|\,\frac{1}{2}, -\frac{1}{2} \right\rangle\left\langle \frac{1}{2},-\frac{1}{2};j + \frac{1}{2},0 \,\Big|\,j,-\frac{1}{2}\right\rangle} \,\hat G^{j+}_{j+1/2, 1/2} \nonumber\\
&\qquad\qquad\qquad+ \textstyle{ \left\langle \frac{1}{2}, \frac{1}{2}; 1, -1 \,\Big|\,\frac{3}{2}, -\frac{1}{2} \right\rangle \left\langle  \frac{3}{2},-\frac{1}{2} ; j + \frac{1}{2},0\,\Big|\,j,-\frac{1}{2} \right\rangle}\,\hat G^{j+}_{j+1/2,3/2}(s)\Big),\\
g'_{j+}(s) &= \sqrt{\frac{2j-2}{2j+1}} \frac{\mpsi}{\Epsi} \, \frac{1}{\mmf{+}}\hat{G}^{j+}_{j-3/2,3/2}(s), \\
f'_{j}(s) &= \sqrt{\frac{2j+2}{2j+1}} \frac{1}{\mmf{+}} \Big(\textstyle{\left\langle \frac{1}{2}, \frac{1}{2}; 1, 0 \,\Big|\,\frac{1}{2}, \frac{1}{2} \right\rangle\left\langle \frac{1}{2},\frac{1}{2};j + \frac{1}{2},0 \,\Big|\,j,\frac{1}{2}\right\rangle} \,\hat G^{j+}_{j+1/2, 1/2} \nonumber\\
&\qquad\qquad\qquad+ \textstyle{ \left\langle \frac{1}{2}, \frac{1}{2}; 1, 0 \,\Big|\,\frac{3}{2}, \frac{1}{2} \right\rangle \left\langle  \frac{3}{2},\frac{1}{2}\, ; j + \frac{1}{2},0\Big|\,j,\frac{1}{2} \right\rangle}\,\frac{1}{\mmf{+}}\hat G^{j+}_{j+1/2,3/2}(s)\Big),\\
g''_{j+}(s) &= \sqrt{\frac{2j-2}{2j+1}} \,\frac{1}{\mmf{+}} \hat{G}^{j+}_{j-3/2,3/2}(s), \\
f''_{j}(s) &= \sqrt{\frac{2j+2}{2j+1}} \frac{1}{\mmf{+}}\textstyle{ \left\langle \frac{3}{2},\frac{3}{2} ; j + \frac{1}{2},0\,\Big|\,j,\frac{3}{2} \right\rangle}\,\hat G^{j+}_{j+1/2,3/2}(s).
\end{align}
\end{subequations}

For the unnatural isobars, the minimal parameterization fulfilling Eq.~\eqref{eq:conspiracy.unnat} is
\begin{subequations}
\label{eq:helicitypw.unnat}
\begin{align}
A_{+,++}^{j-}(s) &= \mmf{-}p^{j-1/2} q^{j-1/2}\bigg[{\textstyle\left\langle \frac{1}{2},\frac{1}{2};1,-1|\frac{1}{2},-\frac{1}{2}\right\rangle\left\langle \frac{1}{2},-\frac{1}{2};j-\frac{1}{2},0|j,-\frac{1}{2}\right\rangle} \, g_{j-}(s) \nonumber\\
&\qquad\qquad\qquad\qquad+ {\textstyle\left\langle \frac{1}{2},\frac{1}{2};1,-1|\frac{3}{2},-\frac{1}{2}\right\rangle\left\langle \frac{3}{2},-\frac{1}{2};j-\frac{1}{2},0|j,-\frac{1}{2}\right\rangle} \, h_{j-}(s) + p^2 \,f_{j+}(s)\bigg],\\
A_{+,+0}^{j-}(s) &= \mmf{-}p^{j-1/2} q^{j-1/2}  \bigg[{\textstyle\left\langle \frac{1}{2},\frac{1}{2};1,0|\frac{1}{2},\frac{1}{2}\right\rangle\left\langle \frac{1}{2},\frac{1}{2};j-\frac{1}{2},0|j,\frac{1}{2}\right\rangle} \, \frac{\Epsi}{\mpsi} \, g'_{j-}(s) \nonumber\\
&\qquad\qquad\qquad\qquad+ {\textstyle\left\langle \frac{1}{2},\frac{1}{2};1,0|\frac{3}{2},\frac{1}{2}\right\rangle\left\langle \frac{3}{2},\frac{1}{2};j-\frac{1}{2},0|j,\frac{1}{2}\right\rangle} \,\frac{\Epsi}{\mpsi} \, h'_{j-}(s) + p^2 \, f_{j-}'(s)\bigg],\\
A_{+,+-}^{j-}(s) &= -\mmf{-} p^{j-1/2} q^{j-1/2}\bigg[{\textstyle\left\langle \frac{3}{2},\frac{3}{2};j-\frac{1}{2},0|j,\frac{3}{2}\right\rangle} \,\left(\frac{1}{C}\, g''_{j-}(s) + h''_{j-}(s)\right) + p^2 \, f_{j-}''(s)\bigg]~,
\end{align}
\end{subequations}
with $g_{j-}(s_\pm) - g'_{j-}(s_\pm) = -g''_{j-}(s_\pm)$ and $h_{j-}(s_\pm) = h'_{j-}(s_\pm) = -h''_{j-}(s_\pm)$. The identification with the \ls couplings is straightforward,
\begin{subequations}
\label{eq:lsmatchingunnat}
\begin{align}
g_{j-}(s) &= \sqrt{\frac{2j}{2j+1}} \, \frac{1}{\mmf{-}} \hat{G}^{j-}_{j-1/2,1/2}(s),
&h_{j-}(s) &= \sqrt{\frac{2j}{2j+1}} \, \frac{1}{\mmf{-}} \hat{G}^{j-}_{j-1/2,3/2}(s), \\
\omit\rlap{$f_{j-}(s) = {\displaystyle\sqrt{\frac{2j+4}{2j+1}}\frac{1}{\mmf{-}}}  \left\langle \frac{1}{2}, \frac{1}{2}; 1, -1 \,\Big|\,\frac{3}{2}, -\frac{1}{2} \right\rangle\left\langle \frac{3}{2},-\frac{1}{2};j + \frac{3}{2},0 \,\Big|\,j,-\frac{1}{2}\right\rangle {\displaystyle\,  \hat G^{j-}_{j+3/2, 3/2}}(s)$,}\\
g'_{j-}(s) &= \sqrt{\frac{2j}{2j+1}} \, \frac{\mpsi}{\Epsi}\, \frac{1}{\mmf{-}} \hat{G}^{j-}_{j-1/2,1/2}(s), &
h'_{j-}(s) &= \sqrt{\frac{2j}{2j+1}} \, \frac{\mpsi}{\Epsi}\,  \frac{1}{\mmf{-}} \hat{G}^{j-}_{j-1/2,3/2}(s), \\
\omit\rlap{$f'_{j-}(s) ={ \displaystyle\sqrt{\frac{2j+4}{2j+1}} \frac{1}{\mmf{-}}} \left\langle \frac{1}{2}, \frac{1}{2}; 1, 0 \,\Big|\,\frac{3}{2}, \frac{1}{2} \right\rangle\left\langle \frac{3}{2},\frac{1}{2};j + \frac{3}{2},0 \,\Big|\,j,\frac{1}{2}\right\rangle {\displaystyle\,\hat G^{j-}_{j+3/2, 3/2}}(s)$,}\\
g''_{j-}(s) &= 0,&
h''_{j-}(s) &= -\sqrt{\frac{2j}{2j+1}} \,\frac{1}{\mmf{-}} \hat{G}^{j-}_{j-1/2,3/2}(s), \\
\omit\rlap{$f''_{j-}(s) =-{ \displaystyle\sqrt{\frac{2j+4}{2j+1}} \frac{1}{\mmf{-}}} \left\langle \frac{1}{2}, \frac{1}{2}; 1, 0 \,\Big|\,\frac{3}{2}, \frac{3}{2} \right\rangle\left\langle \frac{3}{2},\frac{3}{2};j + \frac{3}{2},0 \,\Big|\,j,\frac{3}{2}\right\rangle {\displaystyle\,\hat G^{j-}_{j+3/2, 3/2}}(s)$.}
\end{align}
\end{subequations}
As shown in Eq.~\eqref{eq:lsmatchingnat} and~\eqref{eq:lsmatchingunnat},  care should be taken when choosing a parameterization of the \ls amplitude so that the expressions are free from kinematical singularities, beyond the one at $s=0$ discussed in Section~\ref{sec:macdowell}.. For example, if one  takes
 the functions $\hat G^{j+}_{j-3/2, 3/2}(s)$, $\hat G^{j-}_{j-1/2, 1/2}(s)$ and $\hat G^{j-}_{j-1/2, 3/2}(s)$  to be proportional to
Breit-Wigner functions with constant couplings, the amplitudes $g'_{j+}(s)$, $g'_{j-}(s)$ and $h'_{j-}(s)$ would end up having a pole at $s = \mb^2 - \mpsi^2$.
It is clear that using  Breit-Wigner parameterizations, or any other model for helicity amplitudes, \ie the left-hand sides of Eq.~\eqref{eq:lsmatchingnat} and~\eqref{eq:lsmatchingunnat}, instead of the \ls amplitudes helps prevent unwanted singularities. A practical use of these formulae, and the analogous forms for the PV amplitude and for the $u$-channel exchanges, are given in Appendix~\ref{app:practical}.

\section{MacDowell symmetry}\label{sec:macdowell} 
Up to this point, we have mainly ignored singularities at $s=0$. For the $s$-channel reaction, we choose the scattering configuration such that the pseudothresholds are positive: $\mb - \mpsi > 0$ and $\mpp - \mK > 0$. Our current results therefore depend on the relative sizes of the meson and baryon masses. We then restricted our discussion to positive $\re\sqrt{s}$, meaning that our expressions do not hold for negative $\re\sqrt{s}$. The correct kinematic singularity free amplitudes cannot have this property, since the invariant amplitudes do not depend on relative masses either~\cite{henyey1968threshold}.
The restriction to $\re\sqrt{s} > 0$ limits the reachable kinematic singularities of $\sqrt{\Epp + \mpp}$ and $\sqrt{\Eb + \mb}$ in Eq.~\eqref{eq:eplusm}. For $\re\sqrt{s} < 0$, however, these factors contain (pseudo)threshold branch points.

The above-mentioned complications arise only in the case of fermion-boson scattering, where the total angular momentum is half integer. In this type of process the obtained helicity amplitudes are not invariant under the transformation $\sqrt{s} \to -\sqrt{s}$. It can be argued that for half-integer total angular momenta, the relevant kinematic variable is $\sqrt{s}$ rather than $s$. 

In order to construct a set of amplitudes that is free of kinematical singularities for negative $\re\sqrt{s}$ as well. One must therefore verify that the new (pseudo)treshold singularities at $\re\sqrt{s}<0$ are correctly accounted for by the kinematic factors $K^\eta_{MN}$. We have already silently removed those factors in Eq.~\eqref{eq:KpmMN} by introducing the $Q^\eta$, which take the form
\begin{subequations}
\begin{align}
\mmf{+} &= \left(\sqrt{s} + (\mb + \mpsi)\right)^\frac{1}{2}\left(\sqrt{s} + (\mb - \mpsi)\right)^\frac{1}{2} \left(\sqrt{s} + (\mpp + \mK)\right)^{-\frac{1}{2}}\left(\sqrt{s} + (\mpp - \mK)\right)^{-\frac{1}{2}} \nonumber \\
&= \frac{\sqrt{\Eb + \mb}}{\sqrt{\Epp + \mpp}} \label{eq:Qplus}, \\
\mmf{-} &= \left(\sqrt{s} + (\mb + \mpsi)\right)^{-\frac{1}{2}}\left(\sqrt{s} + (\mb - \mpsi)\right)^{-\frac{1}{2}} \left(\sqrt{s} + (\mpp + \mK)\right)^\frac{1}{2}\left(\sqrt{s} + (\mpp - \mK)\right)^\frac{1}{2} \nonumber \\
&= \frac{\sqrt{\Epp + \mpp}}{\sqrt{\Eb + \mb}} \label{eq:Qminus},\\
\mmf{1/2} &= (\Eb + \mb)\sqrt{s}\,.
\end{align}
\end{subequations}
Under $\sqrt{s} \to -\sqrt{s}$, the terms $p\sqrt{s}$ and $q\sqrt{s}$ remain unchanged. However, the factors $\sqrt{\Epp + \mpp}$ and $\sqrt{\Eb + \mb}$ will now contain branch points at (pseudo)threshold at positive $\re\sqrt{s}$. An additional benefit of this analysis, is that the final amplitudes are now independent of the choice relative masses, i.e.\ the final form of the amplitudes are the same if we would have taken the configuration where the pseudothresholds are negative. 

One can verify that no additional singularities are found at $\re\sqrt{s} < 0$ for $j>\frac{1}{2}$ that have not yet been accounted for. Indeed, this can be done by tracking down the factors of $Q^{\pm}$ in the kinematic matrix $\mathcal{M}^{-1}$ in Eq.~\eqref{eq:grossa}. In other words, the solution to the conspiracy relations for $\re\sqrt{s}>0$ also solve the conspiracy relations for $\re \sqrt{s}<0$. For $j=\frac{1}{2}$ and $\eta = +$, however, no conspiracy relation was needed since the isobars $\hheliso^{j+}(s)$ always appeared with an additional factor $\propto p^2$ for $\re \sqrt{s} > 0$. For $\re \sqrt{s}<0$, however, only $(L=0,S=0)$ is possible in the initial state ($\Lambdab \jpsi$), which requires a new conspiracy relation to be solved. This is indeed reflected by the factor $\mmf{1/2}$ in the definition of $\hheliso^{\frac{1}{2}+}$ in Eq.~\eqref{eq:Ahatdefschannel}. The conspiracy relations read (remember that the $\lambdapsi=-$ contribution vanishes)
\begin{align}
\mpsi \hat A_{+,+ +}^{1/2,+}(s) - (\mb+\sqrt{s})\sqrt{2} \hat A_{+,+ 0}^{1/2,+}(s) &\simeqzero{\Eb \to -\mb} \,, \\
(\mb+\sqrt{s}) \hat A_{+,+ +}^{1/2,+} - \mpsi\sqrt{2} \hat A_{+,+ 0}^{1/2,+} &\simeqzero{\Eb \to -\mb} \,.
\end{align}
Since $\Eb = -\mb$ corresponds to $-\sqrt{s}=\mb+\mpsi$, both relations are fulfilled by requiring that 
\begin{align}
\hat A_{+,+ +}^{1/2,+}(s) + \sqrt{2} \hat A_{+,+ 0}^{1/2,+}(s) \propto  \frac{\hat A_{+,+ +}^{1/2,+}(s)}{ \left\langle \frac{1}{2}, \frac{1}{2}; 1, -1 \,|\,\frac{1}{2}, -\frac{1}{2} \right\rangle} + \frac{\hat A_{+,+ 0}^{1/2,+}(s)}{ \left\langle \frac{1}{2}, \frac{1}{2}; 1, 0 \,|\,\frac{1}{2}, \frac{1}{2} \right\rangle}   &\simeqzero{\Eb \to -\mb}  \,.
\end{align}

Our KSF-PCHA are now free of singularities for both $\re\sqrt{s}> 0$, and $\re\sqrt{s}<0$, and are independent of the chosen mass configuration. Still, they contain remaining singularities at $s=0$. However, MacDowell symmetry~\cite{MacDowell:1959zza} (which is a consequence of CT invariance~\cite{Georgelin:1970sg}) in principle allows one to remove these remaining singularities, resulting in amplitudes that are $s$- and $t$-singularity free~\cite{henyey1968threshold,Arbab:1969zr,Levine:1969ck,Actor:1973cq}. MacDowell symmetry for the KSF-PCHA reads
\begin{align}
F_{\lambdapp,\lambdab \lambdapsi}^{\eta}(-\sqrt{s},t) = \zeta \left(\frac{\mpsi}{\mpp}\right)^{-\eta} F_{\lambdapp,\lambdab \lambdapsi}^{-\eta}(\sqrt{s},t) \,,
\end{align}
where $\zeta = (-1)^{\lambda -\lambda'}$ with  and the factor $(\mpsi/\mpp)^{-\eta}$ is due to our definition of the kinematic factors~$K^\eta_{MN}(s)$ in Eq.~\eqref{eq:KpmMN}. The MacDowell symmetry can be made explicit by considering the relation between the KSF-PCHA and the scalar amplitudes. For example, from Eq.~\eqref{eq:Mmatrix_s_PC} it follows that
\begin{align}
F_{+,+ -}^{\pm}(\sqrt{s},t) \left(\frac{\mpp}{\mpsi}\right)^{\frac{1-\eta}{2}} =- \frac{\mpp^2}{\sqrt{s}}\left[C_2(s,t) + C_4(s,t) (\pm\sqrt{s} - \mb )\right]\,.
\end{align}
In other words, one only needs a single KSF-PCHA for a given helicity combination, say $F_{\lambdapp,\lambdab \lambdapsi}^{+}(\sqrt{s},t)$, and the other one follows from $F_{\lambdapp,\lambdab \lambdapsi}^{+}(-\sqrt{s},t)$. Additionally, the $\mmf{\eta}$ factors defined in Eq.~\eqref{eq:Qplus} introduce a similar behavior for the kinematic factors
\begin{align}
K^\eta_{MN}(-\sqrt{s}) = \left(\frac{\mpsi}{\mpp}\right)^\eta K^{-\eta}_{MN}(\sqrt{s}) .
\end{align}
Considering the definite-parity partial-wave amplitudes in Eq.~\eqref{eq:PCHA}, MacDowell symmetry requires 
\begin{align}\label{eq:weird_equation}
\hheliso^{j\eta}(-\sqrt{s}) &= \zeta \left(\frac{\mpsi}{\mpp}\right)^{-\eta} \hheliso^{j-\eta}(\sqrt{s}). 
\end{align}
This equation might seem odd at first: for a contribution of definite parity, the partial-wave amplitude of definite parity must be non-zero. $N/D$ approaches have been developed, using $\sqrt{s}$ as the relevant variables, rather than $s$~\cite{Gasparyan:2010xz,Gasparyan:2011yw,Lutz:2011xc}. In such a way, the resonances only `resonate' in the definite-parity partial-wave amplitude with the corresponding parity for $\re\sqrt{s}>0$.  
The origin of the singularity that gives rise to the symmetry relation in Eq.~\eqref{eq:weird_equation} can be explained as follows. For unequal masses only, the half-angle factor $\xi_{\lambda\lambda'}(z_s)$ has a branch point at $s=0$. This branch point originates from the factor $\sqrt{1-z_s}^{\abs{\lambda-\lambdap}}$ in Eq.~\eqref{eq:half_angle_factor}.\footnote{Actually, this happens if $(m_1 - m_2)(m_3 - m_4) > 0$, as in our $s$-channel case. If $(m_1 - m_2)(m_3 - m_4) < 0$,    $\sqrt{1-z_s}^{\abs{\lambda-\lambdap}}$ is regular at $s=0$, but the factor $\sqrt{1+z_s}^{\abs{\lambda+\lambdap}}$ in Eq.~\eqref{eq:half_angle_factor} is not. The following discussion proceeds accordingly, leading to the same conclusions.} Hence, taking $\sqrt{s}\to -\sqrt{s}$ results in a phase $\xi_{\lambda\lambda'}(z_s) \to (-1)^{\lambda-\lambda'} \xi_{\lambda\lambda'}(z_s)$. For the second contribution to the PCHA, one has $\xi_{-\lambda\lambda'}(z_s) \to (-1)^{\lambda+\lambda'} \xi_{-\lambda\lambda'}(z_s)$. In particular, the phases $(-1)^{\lambda - \lambda'}$ and $(-1)^{\lambda + \lambda'}$ are the same in all but meson-baryon scattering reactions. Hence, by removing the physical boundary singularities ($z_s = \pm 1$) in forming the $t$-singularity free $d$-functions $\hat d^j_{\lambda,\lambdap}$, we introduced singularities $\sqrt{s}^{\abs{\lambda-\lambda'}}$ as a consequence. For all but meson-baryon scattering reactions, this singularity is removed by dividing the amplitude by $\sqrt{s}^{\abs{\lambda}+\abs{\lambdap}}=\sqrt{s}^{M+N}$. The additional requirement of factorization of the amplitude introduces an extra factor $s^N$, resulting in $\sqrt{s}^{M-N}$ in Eq.~\eqref{eq:KpmMN}.

The remaining kinematical singularities at $s=0$ can now be removed from the KSF-PCHA by exploiting the MacDowell symmetry. Indeed, we can build symmetric and antisymmetric combinations of the $F_{\lambdapp,\lambdab\lambdapsi}^{\eta}(\sqrt{s},t) $, which are even and odd under $\sqrt{s}\to -\sqrt{s}$ respectively, and define totally singularity free functions,
\begin{subequations}
\begin{align}
\hat B_{\lambdapp,\lambdab\lambdapsi}^{+}(s,t) &= F_{\lambdapp,\lambdab\lambdapsi}^{+}(\sqrt{s},t) + \zeta \frac{\mpp}{\mpsi} F_{\lambdapp,\lambdab\lambdapsi}^{-}(\sqrt{s},t), \\
\hat B^{-}_{\lambdapp,\lambdab\lambdapsi}(s,t) &= \frac{1}{\sqrt{s}} \left( F_{\lambdapp,\lambdab\lambdapsi}^{+}(\sqrt{s},t) - \zeta \frac{\mpp}{\mpsi} F_{\lambdapp,\lambdab\lambdapsi}^{-}(\sqrt{s},t) \right),
\end{align}
\end{subequations}
Interestingly enough, these $\hat B_{\lambdapp,\lambdab\lambdapsi}^{\pm}(s,t)$ are free of kinematic singularities in both $s$ and $\sqrt{s}$. 
We remark that~\cite{Anisovich:2004zz} introduces an additional $1/\sqrt{s}$ in the propagator to regularize its high-energy behavior. Such a singular factor cannot be disposed freely, and is incompatible with the MacDowell symmetry.

Despite the fact that we discussed a procedure to remove the singularities at $s = 0$,  Eq.~\eqref{eq:weird_equation} is clearly not compatible with the isobar model. The latter requires isobars with opposite naturalities to be independent, and such constraints cannot be imposed consistently. Although taking care of these singularities is mandatory when considering dispersive analyses, we renounce to do so, for the purpose of making this formalism usable by the isobar practitioners. We therefore set $Q^{\pm,1/2} = 1$ in our final form in Appendix~\ref{app:practical}.

\section{Comparison with the Covariant Projection Method}
 \label{sec:comparison}
The \ls and helicity partial waves can now be compared to the \cpt formalism. The latter builds \ls-like partial-wave amplitudes, based on covariant structures that are interpreted as spin ($S$) and orbital-momentum ($L$) covariant tensors. We follow the methodology outlined by the Bonn-Gatchina partial-wave analysis group in~\cite{Anisovich:2004zz}.
We consider the example of a $\Lambda^*$ resonance with $J^P=\tfrac{3}{2}^-$ in the $s$-channel. First, we consider the interaction in the scattering regime. In this case, all structures must be orthogonalized to the center-of-mass momentum $P=\pb+\ppsi$. We define the relative four-momenta in the intial and final state $p=(\pb - \ppsi)/2$ and $q=(\ppp - \pK)/2$, respectively. The orbital momentum component of the vertex $\Lambda^* \to p \Km$ is described by the $D$-wave tensor
\begin{align}
X^{\rho \mu}(q,P) = \frac{3}{2} q^\rho_\perp q^\mu_\perp- \frac{1}{2} g^{\rho\mu}_\perp q_\perp^2,
\end{align}
with $q_\perp^\mu = q^\mu - P^\mu \,P\cdot q/s$, and $g^{\rho\mu}_\perp = g^{\rho\mu} - P^\rho P^\mu/s$, such that $q_\perp^\mu  P_\mu = q_\rho g_\perp^{\rho \mu} P_\mu = 0$. Furthermore, we define $\gamma_\perp^\mu = g_\perp^{\mu \nu} \gamma_\nu$. The initial state can be $S$-wave or $D$-wave. The orbital tensor structure for the latter reads
\begin{align}
X^{\rho \nu}(p,P) = \frac{3}{2} p^\rho_\perp p^\mu_\perp-\frac{1}{2}g^{\rho\mu}_\perp p_\perp^2,
\end{align}
with $p_\perp^\mu = p^\mu - P^\mu \,P\cdot p/s$. The $\frac{3}{2}^-$ contribution to the helicity amplitudes is therefore fully determined by the expression~\footnote{We neglect overall factors of  $i$.}

\begin{align}\label{eq:cpttensors}
A_{\lambdapp, \lambdab \lambdapsi} = \bar{u}(\ppp,\lambdapp) \gamma_5 \gamma^\perp_\mu  X^{\mu \nu}(q,P) \propagatorP_{\nu\alpha}(P)  &\left[\phantom{+}g_{S_{\frac{3}{2}}}(s) \epsilon^\alpha(\ppsi,\lambdapsi) \right.  \nonumber\\
&\phantom{\Big[}+ g_{D_{\frac{3}{2}}}(s) X^{\alpha \beta}(p,P) \epsilon_\beta (\ppsi,\lambdapsi)\nonumber \\
&\phantom{\Big[}+\left. g_{D_{\frac{1}{2}}}(s) X^{\alpha \beta}(p,P) \gamma^\perp_{\beta} \gamma^\perp_{\delta} \epsilon^\delta (\ppsi,\lambdapsi)
\right] u(\pb,\lambdab),
\end{align}

where we introduced the spin-$\frac{3}{2}$ projector
\begin{align}
\propagatorP_{\mu \nu} = \frac{\slashedP + \sqrt{s}}{2\sqrt{s}} \frac{2}{3} g^\perp_{\mu\alpha} \left( g^\perp - \frac{1}{2}\sigma^\perp \right)^{\alpha \beta} g^\perp_{\beta \nu},
\end{align}
with $\sigma^\perp_{\mu\nu} = \frac{1}{2}\left(\gamma^\perp_\mu \gamma^\perp_\nu - \gamma^\perp_\nu \gamma^\perp_\mu \right)$. Explicitly, the corresponding isobar amplitudes read
\begin{subequations}
\label{eq:cpt_scattering}
\begin{align}
\frac{1}{\pi}A_{+, + +}^{\frac{3}{2}+} &=  \sqrt{\Eb + \mb}\sqrt{\Epp + \mpp} \,q^2 \left[\sqrt{2} g_{S_{\frac{3}{2}}}(s)+ p^2 \left( \frac{1}{\sqrt{2}}  g_{D_{\frac{3}{2}}}(s) + 3\sqrt{2}  g_{D_{\frac{1}{2}}}(s) \right)\right] ,\\
\frac{1}{\pi}A_{+, + 0}^{\frac{3}{2}+} &=  \sqrt{\Eb + \mb}\sqrt{\Epp + \mpp} \, q^2 \frac{\Epsi}{\mpsi} \left[
2 g_{S_{\frac{3}{2}}}(s) + p^2 \left( -2 g_{D_{\frac{3}{2}}}(s) - 3 g_{D_{\frac{1}{2}}}(s) \right) \right], \\
\frac{1}{\pi}A_{+, + -}^{\frac{3}{2}+} &= \sqrt{\Eb + \mb}\sqrt{\Epp + \mpp} \, q^2 \left[ \sqrt{6}g_{S_{\frac{3}{2}}}(s) + p^2 \left( \sqrt{\frac{3}{2} } g_{D_{\frac{3}{2}}}(s) \right)\right]. \label{eq:cpt_scatt_m1}
\end{align}
\end{subequations}
Notice that the expression in Eq.~\eqref{eq:cpt_scatt_m1} indeed does not contain a contribution from the $D_{\frac{1}{2}}$ component, as expected from the \ls in Eq.~\eqref{eq:ls}.
As discussed in Sec.~\ref{sec:s.channel}, the square roots have no singularities at (pseudo)threshold. 
The role of Clebsch-Gordan coefficients can be enlighted by writing 
\begin{align}
 \frac{1}{\pi}A_{+, + \lambdapsi}^{\frac{3}{2}+} &=  \sqrt{\Eb + \mb}\sqrt{\Epp + \mpp} \,q^2 \, \left(\frac{\Epsi}{\mpsi}\right)^{1-|\lambdapsi|} \nonumber \\ 
 &\quad\times\Bigg[ \phantom{+} {\textstyle \left\langle \frac{1}{2},\frac{1}{2};1,-\lambdapsi|\frac{3}{2},\frac{1}{2} - \lambdapsi \right\rangle } \sqrt{6}\, g_{S_{\frac{3}{2}}}(s)\nonumber  \\ 
 &\quad\phantom{\times\Bigg[}+{\textstyle \left\langle 2,0;1, -\lambdapsi | 1,-\lambdapsi \right\rangle \left\langle \frac{1}{2},\frac{1}{2};1,-\lambdapsi|\frac{3}{2},\frac{1}{2} - \lambdapsi \right\rangle } \sqrt{15}\, g_{D_{\frac{3}{2}}}(s) \, p^2 \nonumber \\ 
 &\quad\phantom{\times\Bigg[}-{\textstyle \left\langle \frac{1}{2},\frac{1}{2}-\lambdapsi;2,0 | \frac{3}{2},\frac{1}{2}-\lambdapsi \right\rangle \left\langle \frac{1}{2},\frac{1}{2};1,-\lambdapsi|\frac{1}{2},\frac{1}{2} - \lambdapsi \right\rangle } 3\sqrt{\frac{15}{2}}\, g_{D_{\frac{1}{2}}}(s) \, p^2 \Bigg].
\end{align}
It is worth noticing that the Clebsch-Gordan multiplying the $g_{D_{\frac{3}{2}}}(s)$ coupling is not the one expected according to the \ls construction: the $\jpsi$ spin is coupled with the orbital angular momentum first, and only after with the spin of the the \Lambdab, while the canonical \ls construction would couple the two spins first, and the angular momentum after. This is also evident by looking at the covariant structures the second line of Eq.~\eqref{eq:cpttensors}, and explains why the various tensors are not orthogonal.

The same framework can be applied to the decay chain, where the tensor structures of the initial $\Lambdab \to \jpsi \Lambda^*$ decay must be orthogonalized with the respect to the $\Lambdab$ momentum $\pb$, rather than the isobar momentum $P$. The $\jpsi$ is now in the final state with momentum $\ppsibar = -\ppsi$ and polarization $\epsilon^*_\mu  (\ppsibar, \lambdapsi)$.
We therefore obtain
\begin{align} 
A_{\lambdapp, \lambdab \lambdapsi} = \bar{u}(\ppp,\lambdapp) \gamma_5 \gamma^{\perp}_\mu(P)  X^{\mu \nu}(q,P) \propagatorP_{\nu\alpha}(P)  &\left[\phantom{+}g_{S_{\frac{1}{2}}}(s) 
\epsilon^{\alpha*}(\ppsibar,\lambdapsi) \right. \nonumber \\
&\phantom{\Big[}+ g_{D_{\frac{3}{2}}}(s) X^{\alpha \beta}(p,\pb)
\epsilon^*_\beta (\ppsibar,\lambdapsi)\nonumber \\
&\phantom{\Big[}+\left. g_{D_{\frac{5}{2}}}(s) X^{\alpha \beta}(p,\pb) \gamma^\perp_{\beta}(\pb) \gamma^\perp_{\delta}(\pb)   \epsilon^{*\delta} (\ppsibar,\lambdapsi)
\right] u(\pb,\lambdab).
\end{align}
In the above, the $\gamma^\perp(\pb)$ and $\gamma^\perp(P)$ are orthogonalized with respect to $\pb$ and $P$ respectively. We will show the results for the $g_{D_{\frac{3}{2}}}$ only (equating the other couplings to zero), in the decay chain and scattering regime. In the isobar rest frame, the contribution in the decay chain reads
\begin{subequations}
\label{eq:cpt_decay}
\begin{align}
\frac{1}{\pi}A_{+, + +}^{\frac{3}{2}+} &=  \sqrt{\Eb + \mb}\sqrt{\Epp + \mpp} q^2  p^2 \left[ - \frac{s}{\sqrt{2}  \mb^2} g_{D_{\frac{3}{2}}}(s) \right], \\
\frac{1}{\pi}A_{+, + 0}^{\frac{3}{2}+} &=  \sqrt{\Eb + \mb}\sqrt{\Epp + \mpp} q^2 p^2 \left[  \frac{s\Eb}{\mpsi \mb^4} (s-\mpsi^2 - \mb^2) g_{D_{\frac{3}{2}}}(s) \right], \\
\frac{1}{\pi}A_{+, + -}^{\frac{3}{2}+} &= \sqrt{\Eb + \mb}\sqrt{\Epp + \mpp} q^2 p^2 \left[ -\sqrt{\frac{3}{2}} \frac{s}{\mb^2} g_{D_{\frac{3}{2}}}(s) \right].
\end{align}
\end{subequations}
To summarize, the amplitudes that follow from the \cpt method contain the factor $\Epsi/\mpsi$ in the $\lambdapsi = 0$ isobar. This factor asserts the fulfillment of the conspiracy relation in Eq.~\eqref{eq:conspiracy2.nat}. This factors was included in the canonical helicity amplitudes, but does not follow from the \ls method. 
Additional energy dependent factors $\sqrt{E_p+m_p}\sqrt{E_b+m_b}$ are found, which are not required by analyticity at $\sqrt{s}>0$, since they are smooth. These factors have been discussed in Section~\ref{sec:macdowell}. Even though they are not necessary in the isobar model, one can decide to include them anyways. Since for negative $\re\sqrt{s}$ the minimal orbital angular momentum is given by a $P$-wave in the initial and final state, one expects the kinematic factors $\sqrt{\Eb+\mb} \sqrt{\Epp+\mpp}$, or similarly $\sqrt{\Eb+\mb}q^2/\sqrt{\Epp+\mpp} = \mmf{+} q^2$ to appear. Notice, however, that $\sqrt{E_p+m_p}\sqrt{E_b+m_b} = \mmf{+}(\Epp+\mpp)$, and therefore, a redundant kinematic zero remains at $\Epp=-\mpp$ in Eq.~\eqref{eq:cpt_decay}, which can be reached for negative $\re\sqrt{s}$ only.  As already pointed out in~\cite{Mikhasenko:2017rkh}, the \cpt formalism was shown to violate crossing symmetry, since the amplitudes in the decay and scattering kinematics differ. Also, the coupling of the external particle spins and orbital momentum occur in a different way than in the \ls for the $D_{\frac{3}{2}}$ component. 

\subsection{\texorpdfstring{$p\Km$ mass distribution in different approaches}{pK mass distribution in different approaches}}
We explore the difference between the various approaches and consider two intermediate natural parity, spin-$\frac{3}{2}$ $\Lambdares$ resonances in the $s$-channel ($p \Km$): the $\Lambda(1520)$ with mass $M_{\Lambdares}=1519.5\mev$ and width $\Gamma_{\Lambdares} = 156\mev$ (artificially increased by a factor of $10$ for illustration purposes), and the $\Lambda(1690)$ with $M_{\Lambdares}=1690\mev$ and width $\Gamma_{\Lambdares} = 60\mev$.
We denote  the dynamical part of the amplitude as $T_{\Lambda^*}$.
We consider the \cpt formalism discussed in Eq.~\eqref{eq:cpt_scattering} and Eq.~\eqref{eq:cpt_decay} (for scattering and decay respectively), setting $g_{S_{\frac{3}{2}}}(s) = g_{D_{\frac{1}{2}}}(s) = 0$ and $g_{S_{\frac{1}{2}}}(s) = g_{D_{\frac{5}{2}}}(s) = 0$ respectively. We assume $g_{D_{\frac{3}{2}}}(s) = T_{\Lambdares}(s)$ to be identical in the scattering and decay kinematics, with
\begin{align}
T_{\Lambdares}(s) \equiv \frac{10}{M_{\Lambda(1520)}^2 - s - i M_{\Lambda(1520)} \Gamma_{\Lambda(1520)}} + \frac{1}{M_{\Lambda(1690)}^2 - s - i M_{\Lambda(1690)} \Gamma_{\Lambda(1690)}}.
\end{align}

For the \ls formalism, we choose the couplings in Eq.~\eqref{eq:ls} to be $\hat G^{\frac{3}{2}+}_{0,\frac{3}{2}} = \hat G^{\frac{3}{2}+}_{2,\frac{1}{2}} = 0$ and  $\hat G^{\frac{3}{2}+}_{2,\frac{1}{2}} = T_{\Lambdares}$. The \ls amplitude in the decay kinematics differs from the one in the scattering kinematics only because of the breakup momentum of $\Lambdab \to \jpsi \Lambdares$, calculated in the $\Lambdab$ rest frame or in the $\Lambdares$ rest frame, respectively. Finally, we show the results for our proposed amplitude given in Appendix~\ref{app:practical}. The model is obtained by taking $g_{\Lambda(1520)} + g_{\Lambda(1690)} = T_{\Lambdares}$ in Eq.~\eqref{eq:Asq}.

\begin{figure}
\centering
\includegraphics[width=0.48\textwidth]{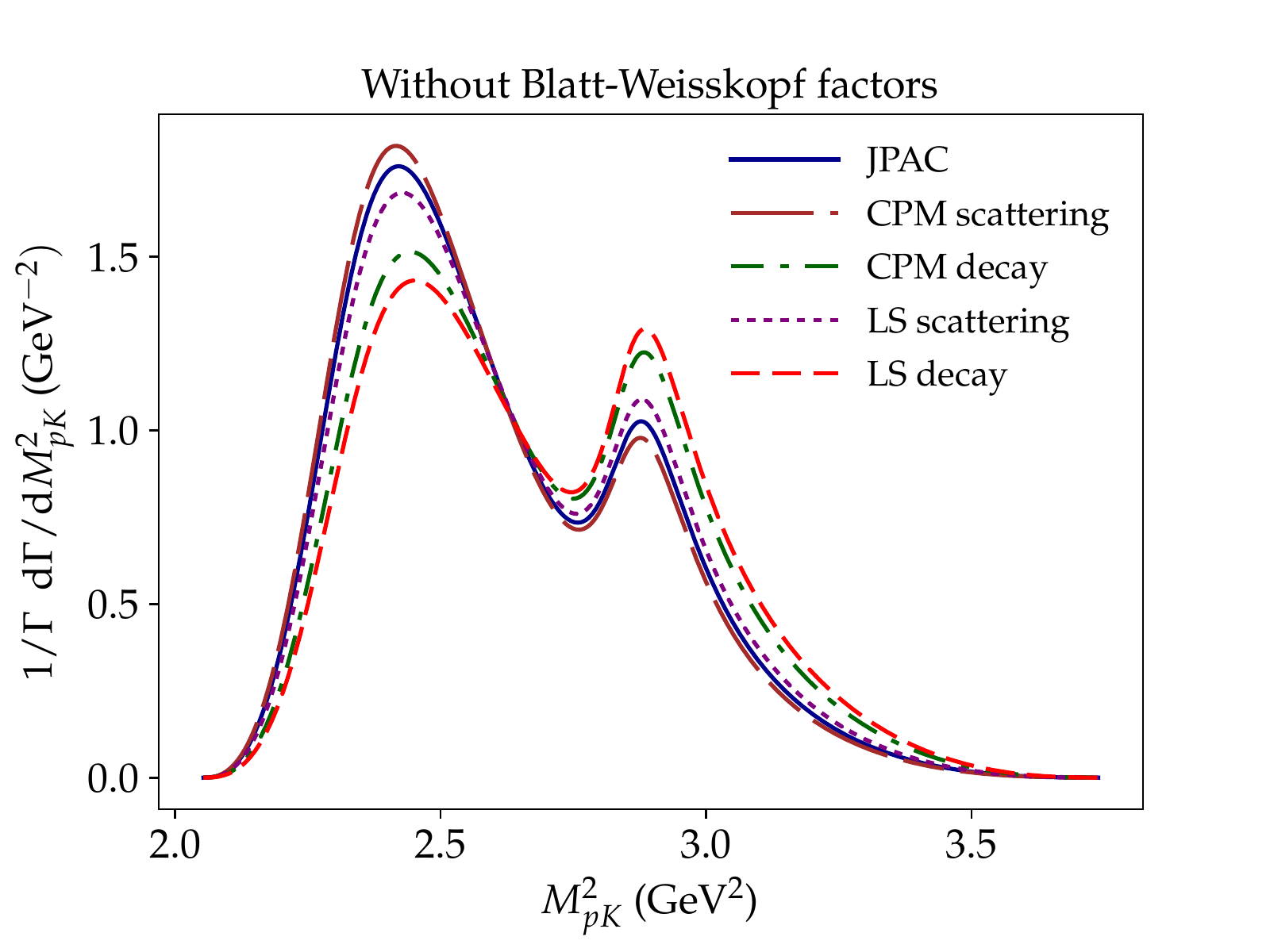}\quad
\includegraphics[width=0.48\textwidth]{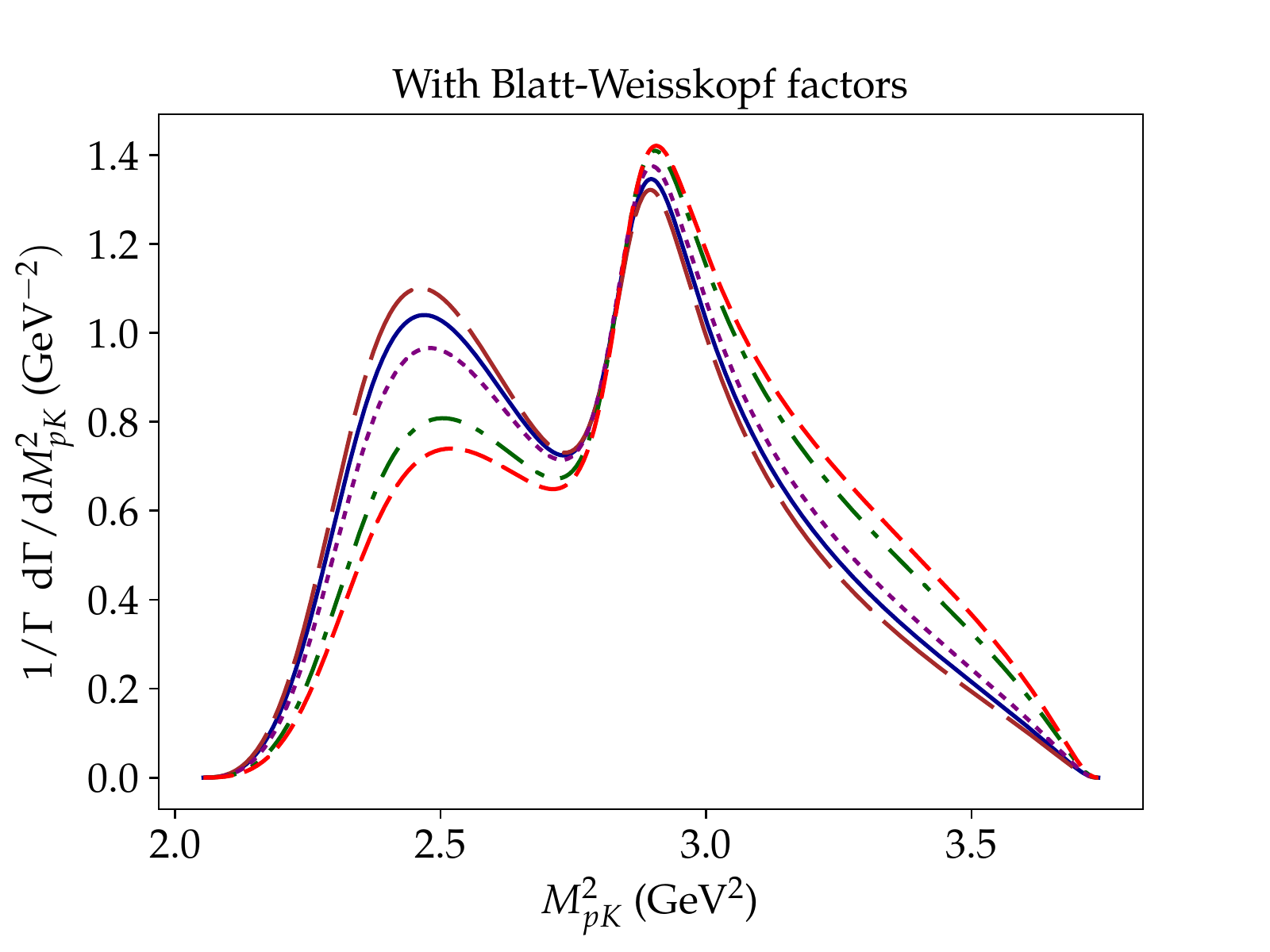}
\caption{Comparison of the line shape of $\Lambda(1520)$ 
(artificially broaden by a factor of 10)
and $\Lambda(1690)$ in the $p\Km$ invariant mass distribution, constructed with the different formalisms. In the left panel we show the result with no barrier factors. In the right panel, we include the customary Blatt-Weisskopf factors.
\label{fig:comparison}}
\end{figure}

As in~\cite{Mikhasenko:2017rkh}, we illustrate the effect of including Blatt-Weisskopf factors in the dynamic part of the amplitude. In the case at hand, this amounts to multiplying the dynamic amplitude $T_{\Lambdares}$ by a factor $B_2(p)B_2(q)$, where $B_2$ is defined as $(x = p,q)$
\begin{align}
  B_2(x) &= \sqrt{\frac{1}{9+3x^2R^2+x^4 R^4}},
\end{align}
and assume $R = 3\gev^{-1}$ as in \cite{Aaij:2015tga}. The differential width is given by
\begin{align}
\frac{\text{d}\Gamma}{\text{d}s} = \sum_j N_j \left( \abs{A^j_{+,++}}^2+\abs{A^j_{+,0+}}^2+\abs{A^j_{+,-+}}^2 \right) \rho(s),
\end{align}
where $\rho(s) = \lambdai^{1/2}\lambdaf^{1/2}/s$ and $N_j$ is a normalization constant. 
The effect of the different kinematic structures is clearly observed in the invariant mass distributions in Fig.~\ref{fig:comparison}. Our proposed amplitudes from 
Appendix~\ref{app:practical}, referred to as the JPAC amplitudes, differ from the \ls amplitudes given in Eq.~\eqref{eq:lsmatchingnat} by the factor of $\Epsi/\mpsi$ in the $\lambdapsi = 0$ helicity partial-wave amplitude ({\it cf.} Eq.~\eqref{eq:our_result}). This factor also follows naturally from the \cpt formalism in the scattering kinematics.
The \cpt amplitudes in the scattering and decay frame (see Eq.~\eqref{eq:cpt_scattering} and Eq.~\eqref{eq:cpt_decay} respectively) both include an additional factor of $\sqrt{\Eb + \mb}\sqrt{\Epp + \mpp}$ compared to the JPAC and \ls formalism, which is related to the discussion in Section~\ref{sec:macdowell}. In addition  the \cpt formalism applied to the decay kinematics introduces redundant kinematic factors of $s$ in all partial-wave amplitudes. Additionally, the $\lambdapsi=0$ amplitude has a factor of $(s-\mb^2-\mpsi^2)\Eb/\Epsi$ in the decay kinematics. 
The differences shown in Fig.~\ref{fig:comparison},
particularly
between the LS decay and the CPM scattering,
are enough to significantly impact the extraction of the couplings. 

\section{Conclusions}
\label{sec:sc}
We have expanded the discussion 
in~\cite{Mikhasenko:2017rkh} 
about the  different approaches for constructing 
amplitudes for scattering and decay 
processes to the fermion-boson case.
In particular, we have studied the 
$\Lambdab\to \psi p \Km$ decay, which is of interest
for hidden charm pentaquark searches.
The inclusion of fermion spins in the helicity formalism
introduces mismatches between threshold and pseudothreshold,
which makes the discussion more complicated and the
equations cumbersome. We used analyticity as 
a guiding principle to examine the canonical helicity
formalism~\cite{Jacob:1959at,Chung:1971ri,Collins:1977jy,cookbook}  and the covariant projection
method~\cite{Chung:1993da,Chung:2007nn,Filippini:1995yc,Anisovich:2006bc}. We have shown how the latter violates
crossing symmetry. The  matching of the helicity amplitudes 
to the most general covariant expression allows us 
to identify the kinematical singularities, 
and to determine the minimal energy dependence required,
summarized in Sec.~\ref{sec:results}. 
In order to factor out the
kinematical singularities we need to build
the hadronic part of the amplitudes with a well defined parity.
A $\sqrt{s}$ singularity cannot be removed with these considerations and needs to be taken care of through the
corresponding dynamical model. As in our previous work~\cite{Mikhasenko:2017rkh}, 
we find meaningful differences among the amplitude building
methodologies  which do affect the resonance pole position extraction, and consequently to the data analysis oriented to
determine the existence and properties of the resonances. This is particularly relevant in situations where several resonances overlaps, and the quantum number assignment is not stable, as in~\cite{Aaij:2015tga}.
Although there is no way to escape all model dependence, 
our analysis maximizes the consistency of 
a given model with the $S$-matrix principles. 
A practical formulation of the amplitudes, 
both in the $s$ and $u$ channels, 
and both for the parity-conserving and parity-violating case,
can be found in Appendix~\ref{app:practical}.

\begin{acknowledgments}
J.N.\ would like to thank Igor Danilkin for bringing Refs.~\cite{Gasparyan:2010xz,Gasparyan:2011yw,Lutz:2011xc} to our attention. This work was supported by BMBF,
the U.S.~Department of Energy under grants No.~DE-AC05-06OR23177 and No.~DE-FG02-87ER40365,
PAPIIT-DGAPA (UNAM, Mexico) grant No.~IA101717,
CONACYT (Mexico) grant No.~251817,
Research Foundation -- Flanders (FWO),
U.S.~National Science Foundation under award numbers PHY-1507572, PHY-1415459 and PHY-1205019,
and Ministerio de Econom\'ia y Competitividad (Spain) through grant
No.~FPA2016-77313-P.
\end{acknowledgments}

\appendix
\section{Polarization vectors and spinors}
\label{app:polarization}
In the $s$-channel center of mass frame the spinors are given by
\begin{subequations}
\label{eq:spinors}
\begin{align}
u\left(\pb,\lambdab = +\tfrac{1}{2}\right) &=
\begin{pmatrix}
\sqrt{\Eb + \mb} \\
0 \\
\sqrt{\Eb - \mb} \\
0
\end{pmatrix}, &
u\left(\pb,\lambdab = -\tfrac{1}{2}\right) &=
\begin{pmatrix}
0 \\
\phantom{-}\sqrt{\Eb + \mb} \\
0 \\
-\sqrt{\Eb - \mb} \\
\end{pmatrix}, \\
u\left(\ppp,\lambdapp = +\tfrac{1}{2}\right) &=
\begin{pmatrix}
\sqrt{\Epp + \mpp} \cos \theta_s/2 \\
\sqrt{\Epp + \mpp} \sin \theta_s/2 \\
\sqrt{\Epp - \mpp} \cos \theta_s/2 \\
\sqrt{\Epp - \mpp} \sin \theta_s/2
\end{pmatrix}, &
u\left(\ppp,\lambdapp = -\tfrac{1}{2}\right) &=
\begin{pmatrix}
-\sqrt{\Epp + \mpp} \sin \theta_s/2 \\
\phantom{-}\sqrt{\Epp + \mpp} \cos \theta_s/2 \\
\phantom{-}\sqrt{\Epp - \mpp} \sin \theta_s/2 \\
-\sqrt{\Epp - \mpp} \cos \theta_s/2
\end{pmatrix} ,
\end{align}
and the $\psi$ polarization by
\begin{align}
\epsilon^\mu(\ppsi,\lambdapsi = \pm 1) &= \frac{1}{\sqrt{2}}\left(0,\pm 1, -i, 0\right), & \epsilon^\mu(\ppsi,\lambdapsi = 0) &= \left(-\frac{p}{\mpsi}, 0, 0, \frac{\Epsi}{\mpsi}\right).
\end{align}
\end{subequations}
We apply the `particle 2' convention for a particle going into the $-z$ direction, as in~\cite{Jacob:1959at}.
The energies $E_i$ are calculated from the momenta and are fully determined by $s$.

\section{\texorpdfstring{Leading and next-to-leading expansion of Wigner $d$-functions}{Leading and next-to-leading expansion of Wigner d-functions}}
\label{app:wigner}
The $\hat d^j_{\lambda\lambda'}(z_s)$ Wigner functions are polynomials in $z_s$ of order $j- M$, with the usual definition of $M = \max(|\lambda|,|\lambda'|)$ and $N = \min(|\lambda|,|\lambda'|)$. 
We use the relation between the Wigner $d$-functions and the Jacobi polynomials $P_n^{(a,b)}$ given by equation (3.74) in~\cite{biedenharn2009angular}
\begin{equation}
\hat{d}_{\lambda\lambdap}^{j}(z_s) = \frac{(-1)^{\frac{1}{2}\left(\left|\lambda - \lambdap\right|+\lambda-\lambdap\right)} }{2^M}
\sqrt{\frac{(j-M)!(j+M)!}{(j-N)!(j+N)!}}\,
P^{(|\lambda-\lambdap|,|\lambda+\lambdap|)}_{j-M}(z_s).
\end{equation}
Two useful relations to compare the above to the literature are $\abs{\lambda + \lambdap} + \abs{\lambda - \lambdap} = 2 M$ and $\abs{\lambda} + \abs{\lambdap} = M+N$. The leading and next-to-leading terms of the polynomial are given by

\begin{align}
\hat d^{j}_{\lambda\lambdap}(z_s ) &= \frac{(-1)^{\frac{1}{2}\left(\left|\lambda - \lambdap\right|+\lambda-\lambdap\right)} }{2^j}\frac{(2j)!}{\sqrt{(j-M)!(j+M)!(j-N)!(j+N)!}}\nonumber\\
&\quad\times\left[z_s^{j-M} +\frac{(j-M)(|\lambda - \lambdap|-M)}{j}z_s^{j-M-1}\right]+ \mathcal{O}(z_s^{j-M-2}).\label{eq:leading_subleading_dhat}
\end{align}
As we noted, this polynomial has no definite parity. 
We defined the parity-conserving Wigner $d$-functions as
\begin{equation}\label{eq:PC.Wigner-d}
\hat d^{j\eta}_{\lambda\lambdap}(z_s) = \hat d^{j}_{\lambda\lambdap}(z_s) + \eta(-1)^{\lambdap - M}\hat d^{j}_{-\lambda\lambdap}(z_s).
\end{equation}
With the substitution $\lambda \to -\lambda$ in Eq.~\eqref{eq:leading_subleading_dhat}, the second term between brackets obtains a minus sign, since $(\abs{\lambda - \lambdap} - M) = -(\abs{\lambda+\lambdap} - M)$, while the first term is unaffected. This illustrates the more general fact that even and odd powers of $z_s$ obtain opposite sign under $\lambda \to -\lambda$. In particular, the leading power in $z_s$ obtains an extra phase $(-1)^{M-\lambda-|\lambda-\lambdap|}$ from the prefactor in front of the brackets. This phase factor is identically equal to $(-1)^{M-\lambdap}$ used in the definition~\ref{eq:PC.Wigner-d}. 
It follows that the $\hat d^{j\eta}_{\lambda\lambdap}(z_s)$ have definite parity, and asymptotic expressions
\begin{align}
\hat d^{j^+}_{\lambda\lambdap}(z_s ) &= \frac{(-1)^{\frac{1}{2}\left(\left|\lambda - \lambdap\right|+\lambda-\lambdap\right)} }{2^{j-1}}\frac{(2j)!}{\sqrt{(j-M)!(j+M)!(j-N)!(j+N)!}} z_s^{j-M}+ \mathcal{O}(z_s^{j-M-2}),\\
\hat d^{j^-}_{\lambda\lambdap}(z_s ) &= \frac{(-1)^{\frac{1}{2}\left(\left|\lambda - \lambdap\right|+\lambda-\lambdap\right)} }{2^{j-1}}\frac{(2j)!(j-M)(|\lambda - \lambdap|-M)}{j\sqrt{(j-M)!(j+M)!(j-N)!(j+N)!}} z_s^{j-M-1}+ \mathcal{O}(z_s^{j-M-3}).
\end{align}

These can be expressed in terms of Clebsch-Gordan coefficients. Explicitly, for the case considered,
\begin{align}
\hat d^{j+}_{-1/2,1/2}(z_s) &\sim \frac{z_s^{j-1/2}f(j)}{\left\langle \frac{1}{2},\frac{1}{2};1,-1|\frac{3}{2},-\frac{1}{2}\right\rangle\left\langle \frac{3}{2},-\frac{1}{2};j-\frac{3}{2},0|j,-\frac{1}{2}\right\rangle}
\sim \frac{z_s^{j-1/2}g(j)}{\left\langle \frac{1}{2},\frac{1}{2};1,-1|\frac{1}{2},-\frac{1}{2}\right\rangle\left\langle \frac{1}{2},-\frac{1}{2};j-\frac{1}{2},0|j,-\frac{1}{2}\right\rangle},\\
 \hat d^{j+}_{1/2,1/2}(z_s) &\sim \frac{z_s^{j-1/2} \sqrt{2} f(j)}{\left\langle \frac{1}{2},\frac{1}{2};1,0|\frac{3}{2},\frac{1}{2}\right\rangle\left\langle \frac{3}{2},\frac{1}{2};j-\frac{3}{2},0|j,\frac{1}{2}\right\rangle}\sim \frac{z_s^{j-1/2}  g(j)}{\sqrt{2}\left\langle \frac{1}{2},\frac{1}{2};1,0|\frac{1}{2},\frac{1}{2}\right\rangle\left\langle \frac{1}{2},\frac{1}{2};j-\frac{1}{2},0|j,\frac{1}{2}\right\rangle},\\
 \hat d^{j+}_{3/2,1/2}(z_s) &\sim \frac{-z_s^{j-3/2} f(j)}{\left\langle \frac{3}{2},\frac{3}{2};j-\frac{3}{2},0|j,\frac{3}{2}\right\rangle}\sim \frac{z_s^{j-3/2} g(j)}{\left\langle \frac{3}{2},\frac{3}{2};j-\frac{1}{2},0|j,\frac{3}{2}\right\rangle}\Bigg(\frac{\left\langle \frac{1}{2},\frac{1}{2};1,-1|\frac{3}{2},-\frac{1}{2}\right\rangle\left\langle \frac{3}{2},-\frac{1}{2};j-\frac{1}{2},0|j,-\frac{1}{2}\right\rangle}{\left\langle \frac{1}{2},\frac{1}{2};1,-1|\frac{1}{2},-\frac{1}{2}\right\rangle\left\langle \frac{1}{2},-\frac{1}{2};j-\frac{1}{2},0|j,-\frac{1}{2}\right\rangle}\nonumber\\
 &\qquad\qquad\qquad\qquad\qquad\qquad\qquad\qquad-\frac{\left\langle \frac{1}{2},\frac{1}{2};1,0|\frac{3}{2},\frac{1}{2}\right\rangle\left\langle \frac{3}{2},\frac{1}{2};j-\frac{1}{2},0|j,\frac{1}{2}\right\rangle}{\left\langle \frac{1}{2},\frac{1}{2};1,0|\frac{1}{2},\frac{1}{2}\right\rangle\left\langle \frac{1}{2},\frac{1}{2};j-\frac{1}{2},0|j,\frac{1}{2}\right\rangle}\Bigg),
\end{align}
where
\begin{align}
f(j)&=\frac{(-1)^{2j+1}(2j)!}{2^{j+3/2}\left(j-\frac{1}{2}\right)! \left(j+\frac{1}{2}\right)!}\sqrt{\frac{4j^2 - 1}{(j-1)j}}, &
g(j) &= \frac{(-1)^{2 j + 1} (2 j)!}{ 2^{j-1/2}\sqrt{3}(j - \tfrac{1}{2})! (j + \tfrac{1}{2})!}\sqrt{
 \frac{2j + 1}{j}},
\end{align}
depends only on $j$.

For the $u$-channel process, we have the initial helicity fixed to $1/2$, and the final one running. We can use the same formulae, upon
\begin{equation}
 \hat d^{j+}_{1/2,\lambda}(z_u) = (-1)^{\lambda-1/2} \hat d^{j+}_{\lambda,1/2}(z_u).
\end{equation}

\section{The matching matrices}
\label{app:matrices}
The matrix $\mathcal{M}$, introduced in Eq.~\eqref{eq:prima} to match the KSF-PCHAs and the covariant basis, is
\begin{equation}
\mathcal{M}=
\begin{pmatrix}
0 & \m{12} & 0 & \m{14} & \m{15} & \m{16}\\
\m{21}  & \m{22} & \m{23} & \m{24} & \m{24} & \m{26}\\
 0 & \m{32} & 0 & \m{34} & 0 & 0\\
 0 &  \m{42}& 0 & \m{44} &  \m{45} & \m{46}\\
 \m{51} &  \m{52}& \m{53} & \m{54} &  \m{55} & \m{56}\\
 0 & \m{62} & 0 & \m{64} & 0 & 0
\end{pmatrix}\label{eq:Mmatrix_s_PC},
\end{equation}
with
\begin{subequations}
\begin{align*}
\m{12} &= \frac{-\left(\Eb - \mb\right) \left(\Epp + \mpp\right) + n(s,t)}{\mpsi}\mpp &
\m{14} &= \frac{ (\Eb - \mb) (\sqrt{s} + \mb)  (\Epp +        \mpp) + (\sqrt{s} - \mb) n(s,t)}{\mpsi}\mpp\\
\m{15} &= \frac{   \Eb - \mb }{\mpsi}2\mpp &
\m{16} &= \frac{  2 (\Eb - \mb) (\sqrt{s} + \mb) \mpp}{\mpsi}\\
\m{21} &=  \frac{ \sqrt{2s} p^2 \mpp}{\mpsi^2}&
\m{22} &=  \frac{\sqrt{2} \mpp (\Epp p^2 + \Epsi n(s,t))}{   \mpsi^2}\\
\m{23} &= \frac{\sqrt{2s} p^2 (\sqrt{s} - \mb) \mpp}{\mpsi^2}&
\m{24} &=\frac{\sqrt{2} (\sqrt{s} - \mb)  (\Epp p^2 +       \Epsi n(s,t))}{\mpsi^2}\mpp\\
\m{25} &=\frac{ \sqrt{2} (\Eb - \mb) (\sqrt{s} + \mb) \mpp}{\mpsi^2}&
\m{26} &=\sqrt{2} (\Eb - \mb)\mpp\\
\m{32} &= -\frac{\mpp^2}{\sqrt{s}}&
\m{34} &=-\frac{\sqrt{s} - \mb }{\sqrt{s}}\mpp^2\\
\m{42} &=  (\Epp + \mpp) \left(\Epp - \mpp - \frac{n(s,t)}{\Eb + \mb}\right)\frac{\mmf{+}}{\mmf{-}}&
\m{44} &=(\Epp +       \mpp) \left((\sqrt{s} - \mb) (\Epp - \mpp) + \frac{\sqrt{s} +          \mb }{\Eb + \mb}n(s,t)\right)\frac{\mmf{+}}{\mmf{-}}\\
\m{45} &=2 (\Epp + \mpp)\frac{\mmf{+}}{\mmf{-}}&
\m{46} &= -2 (\sqrt{s} - \mb) (\Epp + \mpp)\frac{\mmf{+}}{\mmf{-}}\\
\m{51} &= -\frac{\sqrt{2s} (\Eb - \mb) (\Epp + \mpp)}{\mpsi}\frac{\mmf{+}}{\mmf{-}}&
\m{52} &=-\frac{ \sqrt{2} (\Epp + \mpp) (\Epp p^2 +        \Epsi n(s,t))}{(\Eb + \mb) \mpsi}\frac{\mmf{+}}{\mmf{-}}\\
\m{53} &=\frac{ \sqrt{2s}  (\Eb - \mb) (\sqrt{s} + \mb) (\Epp +       \mpp)}{\mpsi}\frac{\mmf{+}}{\mmf{-}}&
\m{54} &=\frac{  \sqrt{2} (\sqrt{s} + \mb) (\Epp +       \mpp) (\Epp p^2 + \Epsi n(s,t))}{(\Eb +       \mb) \mpsi}\frac{\mmf{+}}{\mmf{-}}\\
\m{55} &=\frac{ \sqrt{2} (\sqrt{s} - \mb) (\Epp + \mpp)}{   \mpsi}\frac{\mmf{+}}{\mmf{-}}&
\m{56} &=  -\mpsi \sqrt{2}(\Epp + \mpp)\frac{\mmf{+}}{\mmf{-}}\\
\m{62} &= -\frac{ \mpp (\Epp + \mpp) \mpsi}{(\Eb + \mb) \sqrt{s}} \frac{\mmf{+}}{\mmf{-}}&
\m{64} &= \frac{ (\sqrt{s} + \mb) (\Epp + \mpp)\mpp \mpsi}{(\Eb +        \mb) \sqrt{s}} \frac{\mmf{+}}{\mmf{-}}
\end{align*}
\end{subequations}
The inverse matrix is calculated using {\tt Mathematica}~\cite{Mathematica}. It  is given by
\begin{align}
\mathcal{M}^{-1} &=
\begin{pmatrix}
\mi{11} & \mi{12} & \mi{13} & \mi{14} & \mi{15} & \mi{16}\\
0 &0 & \mi{23} & 0 & 0 & \mi{26}\\
\mi{31} & \mi{32} & \mi{33} & \mi{34} & \mi{35} & \mi{36}\\
0 & 0 & \mi{43} & 0 & 0 & \mi{46}\\
\mi{51} & 0 & \mi{53} & \mi{54} & 0 & \mi{56}\\
\mi{61} & 0 & \mi{63} & \mi{64} & 0 & \mi{66}
\end{pmatrix}
= \frac{1}{p^2} \mathcal{B} + \text{Reg},
\end{align}
with
\begin{align*}
\mi{11} &= -\frac{  \mpsi^3}{4\mpp s p^2}&
\mi{12} &= \frac{ \mpsi^2 \left(\sqrt{s} + \mb\right)}{    2 \sqrt{2} p^2 \mpp s}\\
\omit\rlap{$\displaystyle\mi{13} = \frac{- \left(\Eb + \mb\right) \left[\mpp \mpsi^2 +           \Epp \left(2 \Epsi \mb + \mb^2 - s\right)\right] +       n(s,t) \left(2 \Epsi \mb - \mb^2 + s\right)}{    4 p^2 \mpp^2 \sqrt{s}}$}\\
\mi{14} &= \frac{ \mpsi^2}{  4s   \left(\Epp + \mpp\right) \left(\Eb - \mb\right)} \frac{\mmf{-}}{\mmf{+}} &
\mi{15}& = -\frac{ \mpsi \left(\sqrt{s} - \mb\right)}{   2 \sqrt{2} \left(\Eb - \mb\right) \left(\Epp + \mpp\right) s}\frac{\mmf{-}}{\mmf{+}} \\
\omit\rlap{$\displaystyle\mi{16} = \frac{ n(s,t) \left[-\mb \left(2 \Epsi + \mb\right) + s\right] + \left(\Eb - \mb\right) \left[\mpp \mpsi^2 +          \Epp \left(2 \Epsi \mb - \mb^2 + s\right)\right]}{   4 \left(\Eb - \mb\right) \mpp \left(\Epp + \mpp\right) \mpsi \sqrt{    s}}\frac{\mmf{-}}{\mmf{+}}$}\\
\mi{23} &= -\frac{\mb + \sqrt{s}}{   2  \mpp^2} &
\mi{26} &= -\frac{ \left(\Eb + \mb\right) \left(\sqrt{s} - \mb\right)}{   2 \mpp \left(\Epp + \mpp\right) \mpsi}\frac{\mmf{-}}{\mmf{+}}\\
\mi{31} &= -\frac{ \mpsi \left(\mb + \sqrt{s}\right)}{   4 p^2 \mpp s} &
\mi{32} &= \frac{ \mpsi^2}{  2\mpp s \sqrt{2} p^2} \\
\omit\rlap{$\displaystyle\mi{33} = \frac{\left(\Eb +           \mb\right) \left[\left(\Eb - \mb\right)\left(\Epp + \mpp\right) - \Epsi\left(\Epp - \mpp\right)\right] - \left(\Eb - \Epsi + \mb\right) n(s,t)}{    4 p^2 \mpp^2 \sqrt{s}}$}\\
\mi{34} &= -\frac{ \sqrt{s}-\mb}{   4 \left(\Eb - \mb\right) \left(\Epp + \mpp\right) s}\frac{\mmf{-}}{\mmf{+}}&
\mi{35} &= \frac{ \mpsi}{ 2   \sqrt{2} s \left( \Epp + \mpp\right) \left(\Eb - \mb\right)} \frac{\mmf{-}}{\mmf{+}}\\
\omit\rlap{$\displaystyle\mi{36} =  \frac{ \left(\Eb - \Epsi - \mb\right) n(s,t) - \left(\Eb -           \mb\right) \left[\Epp \left(\Eb - \Epsi + \mb\right) - \mpp \left(\mb + \sqrt{s}\right)\right]}{    4 \left(\Eb - \mb\right) \mpp \left(\Epp + \mpp\right)      \mpsi \sqrt{s}}\frac{\mmf{-}}{\mmf{+}}$}\\
\mi{43} &= -\frac{ 1}{   2 \mpp^2} &
\mi{46} &= \frac{  \left(\Eb + \mb\right)}{ 2 \mpp \mpsi  \left(\Epp +      \mpp\right) }\frac{\mmf{-}}{\mmf{+}}\\
  \mi{51} &=  \frac{\mpsi \left(\sqrt{s}-\mb\right)}{   4 \left(\Eb - \mb\right)  \mpp \sqrt{s}} &
 \mi{53}& = \frac{ n(s,t) \left( \sqrt{s}-\mb \right) + \left(\Eb - \mb\right) \left(\Epp - \mpp\right) \left(\mb +           \sqrt{s}\right)}{4 \left(\Eb - \mb\right) \mpp^2} \\
 \mi{54} &=  \frac{ \mb + \sqrt{s}}{4 \left(\Epp + \mpp\right) \sqrt{s}}\frac{\mmf{-}}{\mmf{+}}&
 \mi{56}& = -\frac{\left(\Eb + \mb\right) \left(\Epp + \mpp\right) \left( \sqrt{s} -\mb \right) +        n(s,t) \left(\mb + \sqrt{s}\right)}{    4  \mpp \left(\Epp + \mpp\right) \mpsi}\frac{\mmf{-}}{\mmf{+}}\\
 \mi{61} &= \frac{\mpsi}{   4 \mpp  \left(\Eb - \mb\right) \sqrt{s}} &
 \mi{63} &= \frac{- \left(\Eb - \mb\right) \left(\Epp - \mpp\right) + n(s,t)}{    4 \left(\Eb - \mb\right) \mpp^2}\\
 \mi{64} &= -\frac{1}{   4 \left(\Epp + \mpp\right) \sqrt{s}} \frac{\mmf{-}}{\mmf{+}}&
 \mi{66} &=  \frac{-\left(\Eb + \mb\right) \left(\Epp + \mpp\right) + n(s,t)}{    4 \mpp \left(\Epp + \mpp\right) \mpsi}\frac{\mmf{-}}{\mmf{+}}
\end{align*}
and the $\mathcal{B}$ matrix introduced in Eq.~\eqref{eq:grossa}, by
\begin{align}
\mathcal{B} &=
\begin{pmatrix}
\B{11} & \B{12} & \B{13} & \B{14} & \B{15} & \B{16}\\
0 &0 & 0 & 0 & 0 & 0\\
\B{31} & \B{32} & \B{33} & \B{34} & \B{35} & \B{36}\\
0 & 0 & 0 & 0 & 0 & 0\\
\B{51} & 0 & \B{53} & 0 & 0 & 0\\
\B{61} & 0 & \B{63} & 0 & 0 & 0
\end{pmatrix},
\label{eq:mostro}\end{align}
with
\begin{align*}
\B{11} &= -\frac{  \mpsi^3}{4\mpp s} &
\B{12} &= \frac{ \mpsi^2 \left(\mb + \sqrt{s}\right)}{    2 \sqrt{2}\mpp s}\\
\omit\rlap{$\displaystyle\B{13} = \frac{ -\left(\Eb + \mb\right) \left(\mpp \mpsi^2 +           \Epp \left(2 \Epsi \mb + \mb^2 - s\right)\right) +       n(s,t) \left(2 \Epsi \mb - \mb^2 + s\right)}{    4 \mpp^2 \sqrt{s}} \equiv \B{13,0}+ n(s,t)\,\B{13,1}$} \\
\B{14} &= \frac{ \mpsi^2\left(\Eb + \mb\right)}{  4s   \left(\Epp + \mpp\right) }\frac{\mmf{-}}{\mmf{+}} \\
\B{15} &= -\frac{ \mpsi \left(\sqrt{s} - \mb\right)\left(\Eb + \mb\right)}{   2 \sqrt{2}  \left(\Epp + \mpp\right) s}\frac{\mmf{-}}{\mmf{+}}  &
\B{16} &= \frac{ n(s,t) \left(-\mb \left(2 \Epsi + \mb\right) + s\right)\left(\Eb + \mb\right)}{   4  \mpp \left(\Epp + \mpp\right) \mpsi \sqrt{    s}}\frac{\mmf{-}}{\mmf{+}}\\
\B{31} &= -\frac{ \mpsi \left(\mb + \sqrt{s}\right)}{   4 \mpp s}  &
\B{32} &= \frac{ \mpsi^2}{  2\mpp s \sqrt{2}} \\
\omit\rlap{$\displaystyle\B{33} = -\frac{\Epsi\left(\Eb +           \mb\right)  \left(\Epp - \mpp\right) + \left(\Eb - \Epsi + \mb\right) n(s,t)}{    4 \mpp^2 \sqrt{s}} \equiv \B{33,0}+ n(s,t)\,\B{33,1}$}\\
\B{34} &= -\frac{\left(\Eb +\mb\right)\left( \sqrt{s}-\mb\right)}{   4  \left(\Epp + \mpp\right) s}\frac{\mmf{-}}{\mmf{+}}\\
\B{35} &= \frac{ \mpsi\left(\Eb + \mb\right)}{ 2   \sqrt{2} s \left( \Epp + \mpp\right) }  \frac{\mmf{-}}{\mmf{+}}&
\B{36} &=  - \frac{ \left(\Eb + \mb\right)\left(-\Eb + \Epsi + \mb\right) n(s,t) }{    4  \mpp \left(\Epp + \mpp\right)      \mpsi \sqrt{s}}\frac{\mmf{-}}{\mmf{+}}\\
 \B{51} &=  \frac{\mpsi \left(\sqrt{s}-\mb\right)\left(\Eb + \mb\right)}{   4   \mpp \sqrt{s}}  &
 \B{53} &= \frac{ n(s,t) \left( \sqrt{s}-\mb \right)\left(\Eb + \mb\right)}{4  \mpp^2} \\
\B{61} &=   \frac{\mpsi\left(\Eb + \mb\right)}{   4 \mpp   \sqrt{s}}  &
\B{63} &= \frac{  n(s,t) \left(\Eb + \mb\right)}{    4  \mpp^2}
\end{align*}

\section{\texorpdfstring{Parity-violating $s$-channel amplitude}{Parity-violating s-channel amplitude}}
\label{app:PV}
The calculation of the PV amplitude is very similar to the PC one carried out in Sec.~\ref{sec:s.channel}. In practice, one effectively needs to consider the $\Lambda_b$ to have $J^P = \tfrac{1}{2}^-$. This turns out into switching the constraints for the natural and unnatural partial waves obtained before. We sketch the derivation. The covariant basis is given by
\begin{subequations}\label{eq:cglnpv}
\begin{align}
M_1^{\prime\mu} &= \pb^\mu, &
M_2^{\prime\mu} &= \ppp^\mu, &
M_3^{\prime\mu} &= \slashedppsi\, \pb^\mu,\\
M_4^{\prime\mu} &= \slashedppsi\, \ppp^\mu, &
M_5^{\prime\mu} &=  \gamma^\mu, &
M_6^{\prime\mu} &= \slashedppsi\, \gamma^\mu,
\end{align}
\end{subequations}
and the kinematical singularity-free helicity partial-wave amplitudes $\hheliso^j(s)$ by
\begin{subequations}
\begin{align}
A^{j\eta}_{\lambdapp, \lambdab \lambdapsi} &= K^\eta_{MN} (pq)^{j-M}\hat A^{j\eta}_{\lambdapp, \lambdab \lambdapsi} & &\text{for }j\ge \tfrac{3}{2},\\
A^{1/2,\eta}_{\lambdapp, \lambdab \lambdapsi} &= \left(\frac{p\sqrt{s}}{\mpsi}\right)^{1-\eta} \left(\mmf{\prime \,1/2}\right)^{\eta} K^\eta_{1/2,1/2} \hat A^{1/2,\eta}_{\lambdapp, \lambdab \lambdapsi}&&\text{for }j={\tfrac{1}{2}} \text{ and }M=\tfrac{1}{2}\label{eq:Kfactorshalfpv},\\
A^{1/2,\eta}_{\lambdapp, \lambdab \lambdapsi} &= 0 &&\text{for }j={\tfrac{1}{2}} \text{ and }M=\tfrac{3}{2},
\end{align}
\end{subequations}
with
\begin{align}
 K^+_{MN} &=\left(\frac{p\sqrt{s}}{\mpsi}\right)^{M - \frac{1}{2}}
\left(\frac{q\sqrt{s}}{\mpp}\right)^{M + \frac{1}{2}} \left(\frac{1}{-\sqrt{s}}\right)^{M - N} \mmf{\prime +}, \\
 K^-_{MN} &=
 \left(\frac{p\sqrt{s}}{\mpsi}\right)^{M - \frac{3}{2}}
\left(\frac{q\sqrt{s}}{\mpp}\right)^{M - \frac{1}{2}} \left(\frac{1}{\sqrt{s}}\right)^{M - N} \mmf{\prime -},
\end{align}
such that $K_{MN}^{-\eta}/K_{MN}^{\eta}=(-)^{M-N}(\tfrac{\mpsi \mpp}{pqs})^\eta \mmf{\prime -\eta}/\mmf{\prime \eta}$. 
The $\mmf{\prime\eta}$ are 
\begin{subequations}
\begin{align}
\mmf{\prime +} &= \frac{1}{\sqrt{s}\sqrt{\Eb + \mb}\sqrt{\Epp + \mpp}}, &
\mmf{\prime -} &= \sqrt{s}\sqrt{\Epp + \mpp}\sqrt{\Eb + \mb},&
\mmf{\prime\,1/2} &= (\Eb + \mb)\sqrt{s}\,
\end{align}
if one considers MacDowell symmetry, as discussed in Section~\ref{sec:macdowell}, or
\begin{equation}
\mmf{\prime\,\pm,1/2} \equiv 1 
\end{equation}
as required by the isobar model and implemented in Appendix~\ref{app:practical}.
\end{subequations}

The matching can be performed in the same way, giving a matching equation analogous to Eq.~\eqref{eq:grossa},
\begin{equation}
\label{eq:grossapv}
 \begin{pmatrix} C_1 \\ C_2 \\ C_3 \\ C_4 \\ C_5 \\ C_6\end{pmatrix}
= \frac{\sqrt{\Epp + \mpp}}{\sqrt{\Eb + \mb}} \mmf{\prime +}\left(\frac{1}{p^2} \mathcal{B}' + \text{Reg}'  \right)\begin{pmatrix}  F^+_{+,++} \\ F^+_{+,+0} \\ F^+_{+,+-} \\ F^-_{+,++} \\ F^-_{+,+0} \\ F^-_{+,+-}\end{pmatrix},
\end{equation}
with the $\mathcal{B}'$ matrix
\begin{align}
\mathcal{B}'=\begin{pmatrix}
D\B{14} & D\B{15} & D\B{16} & \bar{D}\B{11} & \bar{D}\B{12} & \mathcal{R}_{13,0} + \bar{D}\B{13,1}\\
0 &0 & 0 & 0 & 0 & 0\\
D\B{34} & D\B{35} & D\B{36} & \bar{D}\B{31} & \bar{D}\B{32} & \mathcal{R}_{33,0} + \bar{D}\B{33,1}\\
0 & 0 & 0 & 0 & 0 & 0\\
0 & 0 & 0 & \bar{D}\B{51} & 0 & \bar{D}\B{53} \\
0 & 0 & 0 & \bar{D}\B{61} & 0 & \bar{D}\B{63}
\end{pmatrix},
\end{align}
where $D = -\sqrt{s}(1+\Epp/\mpp) \mmf{+}/\mmf{-}$, $\bar D = D^{-1} \mmf{+} \mmf{\prime-}\big/ \mmf{-} \mmf{\prime+}$, the elements of the $\mathcal{B}$ matrix are defined in Appendix~\ref{app:matrices}, and $\B{ij,k}$ stands for the term in $\B{ij}$ of order $[n(s,t)]^k$. Up to irrelevant factors which do not enter the equations (as the multiplicative factors of $D$, or the terms $\mathcal{R}_{13,0} \neq \B{13,0}$ and $\mathcal{R}_{33,0} \neq \B{33,0}$), the conspiracy relations are going to be the same as in Sec.~\ref{sec:s.channel},  upon swapping the natural and unnatural partial waves. For completeness, we report the matrix elements of $\text{Reg}'$:
\begin{equation}
 \text{Reg}'=
\begin{pmatrix}
0 & 0 & \frac{\mpp \mpsi^2 + \Epp \left(-2 \Epsi \mb + \mb^2 - s\right)}{4 \mpp^2 \mpsi} & 0 & 0 & 0\\
0 & 0 & \frac{(\Eb + \mb) (\sqrt{s} - \mb)\sqrt{s}}{2 \mpp^2 \mpsi} & 0 & 0 & -\frac{\sqrt{s} + \mb}{2 \mpp \sqrt{s} (\Epp + \mpp)} \frac{\mmf{\prime-}}{\mmf{\prime+}} \\
0 & 0 & \frac{\Epp (\Eb -\Epsi + \mb) + 
 \mpp (\mb + \sqrt{s})}{4 \mpp^2 \mpsi} & 0 & 0 & 0\\
0 & 0 & -\frac{(\Eb + \mb) \sqrt{s}}{2 \mpp^2 \mpsi} & 0 & 0 & -\frac{1}{2 \sqrt{s} \mpp  (\Epp + \mpp)} \frac{\mmf{\prime -}}{\mmf{\prime +}}\\
-\frac{\mb + \sqrt{s}}{4 \mpp} & 0 & \frac{(\Eb + \mb)(\Epp - \mpp)(\sqrt{s} - \mb) + n(s,t) (\mb + \sqrt{s})}{4 \mpp^2 \mpsi} \sqrt{s} & 0 & 0 & -\frac{\sqrt{s} + \mb}{4\mpp \sqrt{s}}\frac{\mmf{\prime-}}{\mmf{\prime+}}\\
\frac{1}{4 \mpp} & 0 & \frac{(\Eb+\mb)(\Epp - \mpp) - n(s,t)}{4 \mpp^2 \mpsi} \sqrt{s} & 0 & 0 & \frac{1}{4\mpp \sqrt{s}}\frac{\mmf{\prime -}}{\mmf{\prime +}}
\end{pmatrix},
\end{equation}
and
\begin{align}
\mathcal{R}_{13,0} &= (\Eb + \mb)\frac{-\mpp \mpsi^2 + \Epp (2\Epsi \mb + \mb^2 -s)}{4\mpp  (\Epp + \mpp)s} \frac{\mmf{\prime -}}{\mmf{\prime +}} ,  \\
\mathcal{R}_{33,0} &= 
(\Eb + \mb)\frac{\Epp (-\Eb + \Epsi + \mb) + \mpp (\sqrt{s} -\mb)}{4 \mpp (\Epp + \mpp) s} \frac{\mmf{\prime -}}{\mmf{\prime +}}.
\end{align}

\section{\texorpdfstring{The $u$-channel parity-conserving amplitude}{The u-channel parity-conserving amplitude}}
\label{app:uchannel}

\begin{figure}[b]
\centering
\subfigure[\ Diagram]{
\includegraphics[]{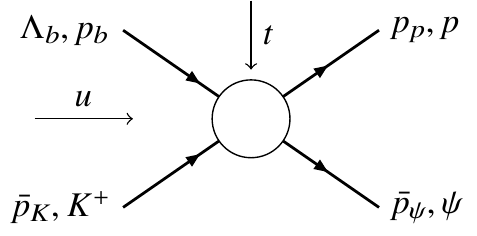}}
\hspace{2cm}
\subfigure[\ Kinematics]{
\includegraphics[]{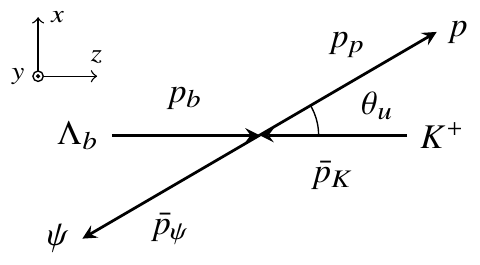}}
\caption{The $u$-channel reaction.} \label{fig:anglesu}
\end{figure}

We briefly review the $u$-channel process $\Lambdab \Kp \to p \psi$ scattering process, where the pentaquark peak is observed. We call $\pbarpsi = -\ppsi$, and $\pbarK = -\pK$ the physical four-momenta of $\psi$ and $\Kp$ in the $u$-channel scattering kinematics. The momentum $\pb$ defines the $z$-axis, the momenta $\ppp$ and $\pbarpsi$ lie in the $xz$-plane, $p_u$ and $q_u$ denote magnitudes of relative momenta in the incoming, $\Lambdab \Kp$ and the outgoing, $p\psi$ states. The scattering angle $\theta_u$ is the polar angle of the proton. The kinematics is summarized in Fig.~\ref{fig:anglesu}. The quantities are expressed
 through the Mandelstam invariants,
\begin{equation}
  \label{eq:zu}
  z_u \equiv \cos \theta_u =\frac{u(t-s)+(\mb^2-\mK^2)(\mpp^2-\mpsi^2)}{4u \,p_u q_u}
  \equiv\frac{n(u,t)}{p_u q_u},\qquad
  p_u = \frac{\lambda_{b K}^{1/2}}{2\sqrt{u}}, \qquad
  q_u = \frac{\lambda_{p \psi}^{1/2}}{2\sqrt{u}}.
\end{equation}
As before, the function $4u\, n(u,t)$ is a polynomial in $u,t$.
For the covariant amplitude, we have
\begin{equation}
\helisou(u,t) = \epsilon^*_\mu(\pbarpsi,\lambdapsi) \, \bar{u}(\ppp,\lambdapp) \left(\sum_{i=1}^6 C_i(u,t)\, M_i^\mu \right) u(\pb,\lambdab),
\end{equation}
with the same covariant basis $M_i$ as in Eq.~\eqref{eq:cgln}; the polarization spinors are the same as in Eq.~\eqref{eq:spinors}, with the obvious replacements $\theta_s \to \theta_u$, $E_{i}(s) \to E_{i}(u)$, with $i=\psi,b,p$. The $\psi$ polarization is given by
\begin{align}
\epsilon^\mu(\pbarpsi,\lambdapsi = \pm 1) &= \frac{1}{\sqrt{2}}\left(0,\pm \cos \theta_u, -i, \mp \sin\theta_u\right), & \epsilon^\mu(\pbarpsi,\lambdapsi = 0) &= \left(-\frac{q_u}{\mpsi},\frac{\Epsi}{\mpsi}\sin\theta_u, 0, \frac{\Epsi}{\mpsi}\cos\theta_u\right).
\end{align}
We remark that, since in the final state the fermion ($p$) is lighter than the boson ($\jpsi$), the factor $\sqrt{\Epp - \mpp}$ will have the threshold singularity only, while $\sqrt{\Epp + \mpp}$ the pseudothreshold singularity only.\footnote{This is, of course, apart from the uncontrolled branch point singularity at $u=0$.} Therefore, the factorized kinematic factors will be different for the threshold and pseudothreshold. The latter is accounted for by the explicit factors of $\sqrt{u-u_-}$ below.
The PCHAs read
\begin{align}
F_{\lambdapp\lambdapsi,\lambdab}^\eta(u,t) & =  \frac{1}{K^\eta_{MN}}\helisou^\eta(u,t) = \frac{1}{4\pi} \sum_{j=3/2} (2j+1)(p_uq_u)^{j-M} \Bigg[ \hhelisou^{j\eta}(u)  \hat d^{j+}_{\lambda \lambda^\prime}(z_u)
+ \hhelisou^{j-\eta}(u) \frac{K^{-\eta}_{MN}}{K^\eta_{MN}}\hat d^{j-}_{\lambda \lambda^\prime}(z_u)\Bigg]\nonumber\\
&\quad + \frac{1}{2\pi} \hhelisou^{1/2,\eta}(u)  \left(\frac{q_u\sqrt{u}}{\mpp}\right)^{1 + \eta} \left(\mmf{1/2}\right)^{-\eta} \sqrt{2}\, (-1)^{\tfrac{1}{2} \left(|\lambda - \lambdap|+\lambda-\lambdap\right)}\delta_{\abs{\lambda'},1/2},
\label{eq:ksf-pchau}
\end{align}
where now $\lambda = \lambdab$, $\lambda' = \lambdapp - \lambdapsi$, and
\begin{align}
 K^+_{MN} &=\left(\frac{p_u\sqrt{u}}{\mb}\right)^{M + \frac{1}{2}}
\left(\frac{q_u\sqrt{u}}{\mpp}\right)^{M - \frac{3}{2}} \sqrt{u - u_-} \left(\frac{1}{\sqrt{u}}\right)^{M - N} \mmf{+}, \\
 K^-_{MN} &=
\left(\frac{p_u\sqrt{u}}{\mb}\right)^{M - \frac{1}{2}}
\left(\frac{q_u\sqrt{u}}{\mpp}\right)^{M - \frac{1}{2}} \frac{1}{\sqrt{u - u_-}} \left(\frac{1}{-\sqrt{u}}\right)^{M - N} \mmf{-},
\end{align}
and $u_\pm = (\mpsi \pm \mpp)^2$ the final-state (pseudo)threshold. 
The $\mmf{\prime\eta}$ are 
\begin{subequations}
\begin{align}
\mmf{+} &= \frac{1}{\sqrt{\Eb+\mb}} \frac{\sqrt{\Epp + \mpp}}{\sqrt{u-u_-}}, &
\mmf{-} &= \sqrt{\Eb+\mb} \frac{\sqrt{u-u_-}}{\sqrt{\Epp + \mpp}} &
\mmf{1/2} &= (\Epp + \mpp)\sqrt{u}
\end{align}
if one considers MacDowell symmetry, as discussed in Section~\ref{sec:macdowell}, or
\begin{equation}
\mmf{\pm,1/2} \equiv 1 
\end{equation}
as required by the isobar model and implemented in Appendix~\ref{app:practical}.
\end{subequations}
Hence, $\tfrac{K_{MN}^{-\eta}}{K_{MN}^{\eta}} = (-)^{M-N}\left(\tfrac{q \mb}{p \mpp (u-u_-)}\right)^{\eta} \frac{\mmf{-\eta}}{\mmf{\eta}}$. The matching equations yield
\begin{equation}
\label{eq:primau}
 \begin{pmatrix} C_1 \\ C_2 \\ C_3 \\ C_4 \\ C_5 \\ C_6\end{pmatrix}
=  \mmf{+}\sqrt{\Eb + \mb} \left(\frac{1}{q_u^2}\mathcal{B} + \frac{1}{u-u_-}\mathcal{P} + \text{Reg}\right)
\begin{pmatrix}  F^+_{++,+} \\ F^+_{+0,+} \\ F^+_{+-,+} \\ F^-_{++,+} \\ F^-_{+0,+} \\ F^-_{+-,+}\end{pmatrix},
\end{equation}
 where the matrices $\mathcal{B}$, $\mathcal{P}$, and $\text{Reg}$ are regular at (pseudo)threshold:
\begin{align}
\mathcal{B} &=
\begin{pmatrix}
0 &  0 & 0  &  0 &  0 & 0\\
\B{21} & \B{22} &  \B{23} &  \B{24} &  \B{25} & \B{26} \\
 0 &  0 & 0  &  0 &  0 &  0 \\
\B{41} & \B{42} & \B{43}  &  \B{44} & \B{45} &  \B{46}\\
 \B{51} &     0 &  \B{53} & 0 &     0 & 0\\
\B{61} &     0 & \B{63} &  0 &     0 & 0
\end{pmatrix}, &
\mathcal{P} &=
\begin{pmatrix}
0 &  0 & 0  &  0 &  0 & 0\\
0 &  0 & 0  &  0 &  0& \PP{26} \\
 0 &  0 & 0  &  0 &  0 &  0 \\
\PP{41} &0 & 0  &  \PP{44} &0 &  \PP{46}\\
0 &0 & 0  &  \PP{54} &0 &  \PP{56}\\
0 &0 & 0  &  \PP{64} &0 &  \PP{66}
\end{pmatrix}, &
\text{Reg} &=
\begin{pmatrix}
0 &  0 & \R{13}  &  0 &  0 & \R{16}\\
0 & 0  & \R{23} & 0 & 0 & \R{26}\\
0 & 0 & \R{33} & 0 & 0 & \R{36}\\
0& 0 & \R{43} & 0 & 0 & \R{46}\\
\R{51} & 0 & \R{53} & 0 & 0 & \R{56}\\
0 & 0 & \R{63} & 0 & 0 & \R{66}
\end{pmatrix},
\end{align}
with
\begin{align*}
\B{21} &= \frac{- \mpp \mpsi^2 \sqrt{u - u_-}}{4 \mb \sqrt{\Epp + \mpp} u} &
\B{22} &= -\frac{ \mpp \mpsi (\mpp + \sqrt{u}) \sqrt{u - u_-}}{2 \sqrt{2} \mb \sqrt{\Epp + \mpp}u }\\
\omit\rlap{$\displaystyle\B{23} = \frac{(\Eb - \mb) (\Epp + \mpp) \mpsi^2 +        n(u,t) \left(\mpsi^2 + 2 (\Epp + \mpp) (\sqrt{u} -\mpp)\right)}{4 \mb^2 \sqrt{\Epp + \mpp} \sqrt{u}}\sqrt{u - u_-}$}\\
\B{24} &= \frac{\sqrt{\Epp + \mpp} \mpsi^2}{4 (\Eb + \mb) u \sqrt{u - u_-}} \frac{\mmf{-}}{\mmf{+}}&
\B{25} &= \frac{\sqrt{\Epp + \mpp} \mpsi (\sqrt{u} -\mpp)}{  2 \sqrt{2} (\Eb + \mb) u \sqrt{u - u_-}}\frac{\mmf{-}}{\mmf{+}} \\
\B{26} &=  \frac{ \sqrt{\Epp + \mpp} (\Epsi + \mpp) \mpsi^2 n(u,t)}{4 \mb (\Eb + \mb) \mpp u \sqrt{u - u_-}}\frac{\mmf{-}}{\mmf{+}}\\
\B{41} &= \frac{\mpp (3 \Epsi \mpp + 2 \mpp^2 + \mpsi^2) \sqrt{u - u_-}}{ 4 \mb \sqrt{\Epp + \mpp} u^{3/2}} &
\B{42} &= \frac{\mpp \mpsi \sqrt{u - u_-}}{2 \sqrt{2} \mb \sqrt{\Epp + \mpp} u}\\
\omit\rlap{$\displaystyle\B{43} = \frac{(\Eb - \mb) (\Epp + \mpp) (\Epsi \mpp +  \mpsi^2) + \left(\Epsi \mpp + \mpp (\Epp + \mpp) - \mpsi^2\right) n(u,t) }{4 \mb^2 \sqrt{\Epp + \mpp}  u}\sqrt{u - u_-}$}\\
\B{44} &= \frac{\sqrt{\Epp + \mpp} (\Epsi \mpp + \mpsi^2)}{4 (\Eb + \mb) u^{3/2} \sqrt{u - u_-}} \frac{\mmf{-}}{\mmf{+}}&
 \B{45} &= \frac{\sqrt{\Epp + \mpp} \mpsi}{2 \sqrt{2} (\Eb + \mb) u \sqrt{u - u_-}}\frac{\mmf{-}}{\mmf{+}} \\
 \B{46} &=  \frac{\sqrt{\Epp + \mpp} (\Epsi \mpp + \mpsi^2) n(u,t)}{4 \mb (\Eb + \mb) \mpp u \sqrt{u - u_-}}\frac{\mmf{-}}{\mmf{+}}&
 \B{51} &= - \frac{\mpp \sqrt{\Epp + \mpp} (\mpsi^2-\Epsi \mpp) \sqrt{u - u_-}}{4 \mb (\mpsi^2-\mpp^2 ) }\\
 \B{53} &= -\frac{\sqrt{\Epp + \mpp} n(u,t) ( \sqrt{u}-\mpp) \sqrt{u - u_-}}{4 \mb^2}&
 \B{61} &= -\frac{ \mpp \sqrt{\Epp + \mpp} \sqrt{u - u_-}}{4 \mb  \sqrt{u}}\\
 \B{63} &=  - \frac{\sqrt{\Epp + \mpp} n(u,t) \sqrt{u - u_-}}{4 \mb^2 },
\end{align*}
\begin{align*}
\omit\rlap{$ \PP{26} = \frac{\textstyle n(u,t) \left(2\Epp\Epsi + 3 \mpsi^2\right)+ \Eb (\Epsi - \mpp) \mpsi^2  + \mb \mpsi^2 \sqrt{u} }{\textstyle 4 \mb (\Eb + \mb) \mpp  u } \frac{\textstyle \sqrt{u - u_-}}{\textstyle \sqrt{ \Epp + \mpp}}\frac{\textstyle\mmf{-}}{\textstyle\mmf{+}}$}\\
\PP{41} &= \frac{\mpp \left(2 \sqrt{u} + 3 \mpp\right) }{4 \mb  u^{3/2}} \frac{(u - u_-)^{3/2}}{(\Epp + \mpp)^{3/2}} &
\PP{44} &= \frac{\mpp + 2 \sqrt{u}}{4 (\Eb + \mb) u^{3/2}}\frac{ \sqrt{u - u_-}}{ \sqrt{\Epp + \mpp}}\frac{\mmf{-}}{\mmf{+}}\\
\PP{46} &= \frac{(\Eb +  \mb) (\Epsi \mpp - \mpsi^2) + \mpp n(u,t)}{4 \mb (\Eb + \mb) \mpp \sqrt{\Epp + \mpp} u }\sqrt{u - u_-}\frac{\mmf{-}}{\mmf{+}}\\
\PP{54} &=  -\frac{\mpp + \sqrt{u}}{4 (\Eb + \mb)  \sqrt{u}}\frac{\sqrt{u - u_-}}{\sqrt{\Epp + \mpp}}\frac{\mmf{-}}{\mmf{+}}&
\PP{56} &= \frac{n(u,t) (\mpp + \sqrt{u})\sqrt{u - u_-}}{4 \mb (\Eb + \mb) \mpp \sqrt{\Epp + \mpp} }\frac{\mmf{-}}{\mmf{+}}\\
\PP{64} &= \frac{\sqrt{u - u_-}}{4\sqrt{\Epp + \mpp}  (\Eb + \mb)  \sqrt{u} }\frac{\mmf{-}}{\mmf{+}}&
\PP{66} &=  -\frac{n(u,t) \sqrt{u - u_-}}{4 \mb (\Eb + \mb) \mpp \sqrt{\Epp + \mpp} }\frac{\mmf{-}}{\mmf{+}},
\end{align*}
and
\begin{align*}
 \R{13} &= -\frac{(\sqrt{u} + \mpp) \sqrt{u - u_-}}{ 2 \mb^2 \sqrt{\Epp + \mpp}} &
 \R{16} &= -\frac{ \sqrt{\Epp + \mpp} (\sqrt{u} - \mpp)}{ 2 \mb (\Eb + \mb) \mpp \sqrt{u - u_-}}\frac{\mmf{-}}{\mmf{+}}\\
 \R{23} &= \frac{\Eb (\mpp + \sqrt{u}) \sqrt{u - u_-}}{ 2 \mb^2 \sqrt{\Epp + \mpp} \sqrt{u}}\\
\omit\rlap{$ \R{26} = \frac{ {\textstyle 2n(u,t) (\Epp + \sqrt{u} - \mpp)+ \Eb ( 2u+  \mpsi^2 - 2 \mpp \sqrt{u})}}{\textstyle 4 \mb (\Eb + \mb) \mpp  u }\frac{\textstyle\sqrt{ \Epp + \mpp}}{\textstyle\sqrt{u - u_-}}\frac{\textstyle\mmf{-}}{\textstyle\mmf{+}}$}\\
\R{33} &= \frac{\sqrt{u - u_-}}{2 \mb^2 \sqrt{\Epp + \mpp}} &
\R{36} &= -\frac{\sqrt{\Epp + \mpp}}{ 2 \mb (\Eb + \mb) \mpp \sqrt{u - u_-}}\frac{\mmf{-}}{\mmf{+}}\\
\R{43} &= \frac{ \mpp(\Eb-\mb) - 2\mb  \sqrt{u}}{4 \mb^2 \sqrt{\Epp + \mpp} u} \sqrt{u - u_-} &
\R{46} &= -\frac{\sqrt{\Epp + \mpp}  \left(2\mb \sqrt{u}- (\Eb + \mb) \mpp\right)}{4 \mb (\Eb + \mb) \mpp  \sqrt{u - u_-}}\frac{\mmf{-}}{\mmf{+}}\\
\R{51} &= -\frac{\mpp^2 \sqrt{u - u_-}}{4 \mb \sqrt{\Epp + \mpp} (\mpsi^2 - \mpp^2)} &
\R{53} &= - \frac{(\Eb - \mb) (\mpp + \sqrt{u}) \sqrt{u - u_-}}{ 4 \mb^2 \sqrt{\Epp + \mpp}}\\
\R{56} &= \frac{ (\sqrt{u} -\mpp)\sqrt{\Epp + \mpp}}{4 \mb \mpp  \sqrt{u - u_-}} \frac{\mmf{-}}{\mmf{+}}&
\R{63} &= \frac{(\Eb - \mb) \sqrt{u - u_-}}{4 \mb^2 \sqrt{\Epp + \mpp}} \\
\R{66} &=  \frac{\sqrt{\Epp + \mpp}}{4 \mb \mpp \sqrt{u - u_-}}\frac{\mmf{-}}{\mmf{+}}.
\end{align*}
Notice that the factor $\sqrt{(u-u_-)/(E_p+m_p)}$ is regular at pseudothreshold. Since the set of conspiracy equations does not have a nontrivial solution for general $u$, we follow the same argument we used for the unnatural isobars in Sec.~\ref{sec:s.channel}. Imposing the constraints at threshold for the $\hhelisou$, and restoring the kinematic factors, we get the conspiracy relations expected from the \ls analysis, analogous to the ones in Eqs.~\eqref{eq:conspiracy1.nat},~\eqref{eq:conspiracy2.nat},  and~\eqref{eq:conspiracy.unnat}
\begin{subequations}
 \label{eq:conspiracy.u.thr}
\begin{align}
&\frac{A_{++,+}^{j+}(u_+) }{\left\langle \frac{1}{2},\frac{1}{2};1,-1|\frac{3}{2},-\frac{1}{2}\right\rangle\left\langle \frac{3}{2},-\frac{1}{2};j-\frac{3}{2},0|j,-\frac{1}{2}\right\rangle}
- \frac{  A_{+-,+}^{j+}(u_+) }{\left\langle \frac{3}{2},\frac{3}{2};j-\frac{3}{2},0|j,\frac{3}{2}\right\rangle} \simeqzero{u-u_+} ,\\
&\frac{A_{+0,+}^{j+}(u_+)}{\left\langle \frac{1}{2},\frac{1}{2};1,0|\frac{3}{2},\frac{1}{2}\right\rangle\left\langle \frac{3}{2},\frac{1}{2};j-\frac{3}{2},0|j,\frac{1}{2}\right\rangle}  - \frac{ A_{++,+}^{j+}(u_+)}{\left\langle \frac{1}{2},\frac{1}{2};1,-1|\frac{3}{2},-\frac{1}{2}\right\rangle\left\langle \frac{3}{2},-\frac{1}{2};j-\frac{3}{2},0|j,-\frac{1}{2}\right\rangle}\simeqzero{u-u_+},\\
&\frac{A^{j-}_{++,+}(u_+)}{\left\langle \frac{1}{2},\frac{1}{2};1,-1|\frac{1}{2},-\frac{1}{2}\right\rangle\left\langle \frac{1}{2},-\frac{1}{2};j-\frac{1}{2},0|j,-\frac{1}{2}\right\rangle} -
 \frac{A^{j-}_{+0,+}(u_+) }{\left\langle \frac{1}{2},\frac{1}{2};1,0|\frac{1}{2},\frac{1}{2}\right\rangle\left\langle \frac{1}{2},\frac{1}{2};j-\frac{1}{2},0|j,\frac{1}{2}\right\rangle}- \frac{ A^{j-}_{+-,+}(u_+) \,C}{\left\langle \frac{3}{2},\frac{3}{2};j-\frac{1}{2},0|j,\frac{3}{2}\right\rangle} \simeqzero{u-u_+}, \label{eq:uchanunnatm}
\end{align}
\end{subequations}
where $C$ was defined in Eq.~\eqref{eq:C.clebsches}. The $\simeqzero{u-u_+}$ indicates that, for $u \to u_+$, the left hand part of the equation vanishes as fast as $q_u^{j+ 1 - \eta/2}$. These equations agree with the predictions for the \ls couplings. The same conspiracy equations hold at pseudothreshold for the natural case, whereas for the unnatural isobars at threshold we find
\begin{subequations}
 \label{eq:conspiracy.u.ps}
\begin{align}
&\frac{A^{j-}_{++,+}(u_-)}{\left\langle \frac{1}{2},\frac{1}{2};1,-1|\frac{1}{2},-\frac{1}{2}\right\rangle\left\langle \frac{1}{2},-\frac{1}{2};j-\frac{1}{2},0|j,-\frac{1}{2}\right\rangle} + \frac{A^{j-}_{+-,+}(u_-) \,C}{\left\langle \frac{3}{2},\frac{3}{2};j-\frac{1}{2},0|j,\frac{3}{2}\right\rangle}\simeqzero{u-u_-}, \\
 &\frac{A^{j-}_{+0,+}(u_-) }{\left\langle \frac{1}{2},\frac{1}{2};1,0|\frac{1}{2},\frac{1}{2}\right\rangle\left\langle \frac{1}{2},\frac{1}{2};j-\frac{1}{2},0|j,\frac{1}{2}\right\rangle} + 2 \frac{A^{j-}_{+-,+}(u_-) \,C}{\left\langle \frac{3}{2},\frac{3}{2};j-\frac{1}{2},0|j,\frac{3}{2}\right\rangle} \simeqzero{u-u_-}.
\end{align}
\end{subequations}
The latter system of equations overconstrains the relations between the unnatural isobars, which are expected from the \ls in Eq.~\eqref{eq:uchanunnatm} to depend on two independent functions. To avoid this, we impose all these functions to vanish independently at pseudothreshold.
All these constraints are satisfied by choosing
\begin{subequations}
\label{eq:helicitypw.u}
\begin{align}
A_{++,+}^{j+}(u) &= \mmf{+}p_u^{j+1/2} q_u^{j-3/2}\bigg[{\textstyle\left\langle \frac{1}{2},\frac{1}{2};1,-1|\frac{3}{2},-\frac{1}{2}\right\rangle\left\langle \frac{3}{2},-\frac{1}{2};j-\frac{3}{2},0|j,-\frac{1}{2}\right\rangle} \, g_{j+}(u) + q_u^2 \,f_{j+}(u)\bigg],\\
A_{+0,+}^{j+}(u) &= \mmf{+}p_u^{j+1/2} q_u^{j-3/2}\bigg[{\textstyle\left\langle \frac{1}{2},\frac{1}{2};1,0|\frac{3}{2},\frac{1}{2}\right\rangle\left\langle \frac{3}{2},\frac{1}{2};j-\frac{3}{2},0|j,\frac{1}{2}\right\rangle} \,   g'_{j+}(u) + q_u^2 \, f_{j+}'(s)\bigg],\\
A_{+-,+}^{j+}(u) &= \mmf{+}p_u^{j+1/2} q_u^{j-3/2}\bigg[{\textstyle\left\langle \frac{3}{2},\frac{3}{2};j-\frac{3}{2},0|j,\frac{3}{2}\right\rangle} g''_{j+}(u) + q_u^2 \, f_{j+}''(u)\bigg],\\
A_{++,+}^{j-}(u) &= \mmf{-}p_u^{j-1/2} q_u^{j-1/2}\bigg[\frac{\Epp+\mpp}{2\mpp}{\textstyle  \Big(\left\langle \frac{1}{2},\frac{1}{2};1,-1|\frac{1}{2},-\frac{1}{2}\right\rangle\left\langle \frac{1}{2},-\frac{1}{2};j-\frac{1}{2},0|j,-\frac{1}{2}\right\rangle} \, g_{j-}(u) \nonumber\\
&\qquad\qquad\qquad\qquad+ {\textstyle\left\langle \frac{1}{2},\frac{1}{2};1,-1|\frac{3}{2},-\frac{1}{2}\right\rangle\left\langle \frac{3}{2},-\frac{1}{2};j-\frac{1}{2},0|j,-\frac{1}{2}\right\rangle} \, h_{j-}(u)\Big) + q_u^2 \,f_{j+}(u)\bigg],\\
A_{+0,+}^{j-}(u) &= \mmf{-}p_u^{j-1/2} q_u^{j-1/2}\bigg[\frac{\Epp+\mpp}{2\mpp}\Big({\textstyle\left\langle \frac{1}{2},\frac{1}{2};1,0|\frac{1}{2},\frac{1}{2}\right\rangle\left\langle \frac{1}{2},\frac{1}{2};j-\frac{1}{2},0|j,\frac{1}{2}\right\rangle} \,  g'_{j-}(u) \nonumber\\
&\qquad\qquad\qquad\qquad+ {\textstyle\left\langle \frac{1}{2},\frac{1}{2};1,0|\frac{3}{2},\frac{1}{2}\right\rangle\left\langle \frac{3}{2},\frac{1}{2};j-\frac{1}{2},0|j,\frac{1}{2}\right\rangle}\, h'_{j-}(u)\Big) + q_u^2 \, f_{j-}'(u)\bigg],\\
A_{+-,+}^{j-}(u) &= -\mmf{-}p_u^{j-1/2} q_u^{j-1/2}\bigg[\frac{\Epp+\mpp}{2\mpp}{\textstyle\left\langle \frac{3}{2},\frac{3}{2};j-\frac{1}{2},0|j,\frac{3}{2}\right\rangle} \,\left(\frac{1}{C}\, g''_{j-}(u) + h''_{j-}(u)\right) + q_u^2 \, f_{j-}''(u)\bigg],
\end{align}
\end{subequations}
with $g_{j+}(u_\pm) = g'_{j+}(u_\pm) = g''_{j+}(u_\pm)$, $g_{j-}(u_\pm) - g'_{j-}(u_\pm) = g''_{j-}(u_\pm)$ and $h_{j-}(u_\pm) = h'_{j-}(u_\pm) = h''_{j-}(u_\pm)$. 
Note that the choice of the factor $\Epp+\mpp$ in the $A_{+\lambdapsi,+}^{j-}(u)$ can be also justified through the singularity analysis for negative $\sqrt{u}$ as well, as in Section~\ref{sec:macdowell}.

\section{\texorpdfstring{The $u$-channel parity-violating amplitude}{The u-channel parity-violating amplitude}}
\label{app:uchannelpv}
To carry out the analysis for the PV part of the $u$-channel amplitude, we remark that changing the spin-parity of the $\Lambdab$ from  $J^P = \dfrac{1}{2}^+$ to $J^P = \dfrac{1}{2}^-$ only affects the arguments related to the initial state $\Lambdab \Kp$. Since the arguments in the previous channel were based on the (pseudo)threshold of the final state $p\jpsi$, the derivation of the conspiracy relations is unaffected. Therefore, the kinematical factors are identical to the ones for the PC $u$-channel amplitudes, and will not be discussed any further. For completeness, we report the matching matrices,
\begin{equation}
 \begin{pmatrix} C_1 \\ C_2 \\ C_3 \\ C_4 \\ C_5 \\ C_6\end{pmatrix}
=  \mmf{\prime +}\sqrt{\Eb + \mb} \left(\frac{1}{q_u^2}\mathcal{B}' + \frac{1}{u-u_-}\mathcal{P}' + \text{Reg}'\right)
\begin{pmatrix}  F^+_{++,+} \\ F^+_{+0,+} \\ F^+_{+-,+} \\ F^-_{++,+} \\ F^-_{+0,+} \\ F^-_{+-,+}\end{pmatrix},
\end{equation}
The $\mmf{\prime\eta}$ are 
\begin{subequations}
\begin{align}
\mmf{\prime+} &= \sqrt{u} \sqrt{\Eb + \mb} \frac{\sqrt{\Epp+\mpp}}{\sqrt{u-u_-}} , &
\mmf{\prime-} &= \frac{1}{ \sqrt{u} \sqrt{\Eb + \mb}} \frac{\sqrt{u-u_-}}{\sqrt{\Epp+\mpp}} &
\mmf{\prime\,1/2} &= (\Epp + \mpp)\sqrt{u}
\end{align}
if one considers MacDowell symmetry, as discussed in Section~\ref{sec:macdowell}, or
\begin{equation}
\mmf{\prime\,\pm,1/2} \equiv 1 
\end{equation}
as required by the isobar model and implemented in Appendix~\ref{app:practical}.
\end{subequations}
The matrices are given by
\begin{align}
\mathcal{B}' &=
\begin{pmatrix}
0 & 0 & 0 & 0 & 0 & 0 \\
\Bp{21} & \Bp{22} & \Bp{23} & \Bp{24} & \Bp{25} & \Bp{26}\\
0 & 0 & 0 & 0 & 0 & 0 \\
\Bp{41} & \Bp{42} & \Bp{43} & \Bp{44} & \Bp{45} & \B{46}\\
\Bp{51} & 0 & \Bp{53} & 0 & 0 & 0\\
\Bp{61} & 0 & \Bp{63} & 0 & 0 & 0
\end{pmatrix},
& \mathcal{P}' &=
\begin{pmatrix}
0 & 0 & 0 & 0 & 0 & 0 \\
0 & 0 & 0 & 0 & 0 & \Pp{26}\\
0 & 0 & 0 & 0 & 0 & 0 \\
0 & 0 & 0 & 0 & 0 & \Pp{46}\\
0 & 0 & 0 & \Pp{54} & 0 & \Pp{56}\\
0 & 0 & 0 & \Pp{64} & 0 & \Pp{66}
\end{pmatrix},
& \text{Reg}' &=
\begin{pmatrix}
0 & 0 & \Rp{13} & 0 & 0 & \Rp{16} \\
0 & 0 & \Rp{23} & 0 & 0 & 0 \\
0 & 0 & \Rp{33} & 0 & 0 & \Rp{36}\\
0 & 0 & \Rp{43} & 0 & 0 & \Rp{46}\\
0 & 0 & \Rp{53} & 0 & 0 & \Rp{56}\\
0 & 0 & \Rp{63} & 0 & 0 & \Rp{66}
\end{pmatrix},
\end{align}
where
\begin{align*}
\Bp{21} &= -\frac{\mpp \mpsi^2 \sqrt{u - u_-}}{4 (\Eb + \mb) \sqrt{\Epp + \mpp}  u^{3/2}} &
\Bp{22} &= -\frac{\mpp \mpsi (\mpp + \sqrt{u}) \sqrt{u - u_-}}{2 \sqrt{2} (\Eb + \mb) \sqrt{\Epp + \mpp} u^{3/2}}\\
\omit\rlap{$\Bp{23} = \frac{\textstyle\Eb (\Epp + \mpp) \mpsi^2 + \mb (\Epp + \mpp) \mpsi^2 + \left(2 \Epsi (\Epp + \mpp) + \mpsi^2\right) n(u,t)}{\textstyle 4 \mb (\Eb + \mb) \sqrt{\Epp + \mpp}  u}\sqrt{u - u_-}$}\\
\Bp{24} &= \frac{\sqrt{\Epp + \mpp} \mpsi^2}{4 \mb  \sqrt{u} \sqrt{u - u_-}} \frac{\mmf{\prime -}}{\mmf{\prime +}}&
\Bp{25} &= \frac{\sqrt{\Epp + \mpp} \mpsi (\sqrt{u} - \mpp)}{ 2 \sqrt{2} \mb \sqrt{u} \sqrt{u - u_-}}\frac{\mmf{\prime -}}{\mmf{\prime +}}\\
\Bp{26} &= -\frac{\sqrt{\Epp + \mpp} n(u,t) (2 \Epsi \mpp + \mpp^2 - s)}{4 \mb^2 \mpp \sqrt{u - u_-}}\frac{\mmf{\prime -}}{\mmf{\prime +}}&
\Bp{41} &= -\frac{ \mpp (\mpp + \sqrt{u}) \sqrt{u - u_-}}{ 4 (\Eb + \mb) \sqrt{\Epp + \mpp} u^{3/2}}\\
\Bp{42} &= -\frac{\mpp \mpsi \sqrt{u - u_-}}{2 \sqrt{2} (\Eb + \mb) \sqrt{\Epp + \mpp} u^{3/2}}\\
\omit\rlap{$ \Bp{43} = -\frac{\textstyle\mb (\Epsi - \mpp) \mpp + \Eb \Epsi (\Epp + \mpp) + (\Epp + \mpp - \Epsi ) n(u,t) + \Epp \mb \sqrt{u}}{\textstyle 4 \mb (\Eb + \mb) \sqrt{\Epp + \mpp}  u} \sqrt{u - u_-}$}\\
\Bp{44} &= -\frac{\sqrt{\Epp + \mpp} (\sqrt{u} - \mpp)}{4 \mb \sqrt{u} \sqrt{u - u_-}}\frac{\mmf{\prime -}}{\mmf{\prime +}}&
\Bp{45} &= -\frac{\sqrt{\Epp + \mpp} \mpsi}{ 2 \sqrt{2} \mb \sqrt{u} \sqrt{u - u_-}}\frac{\mmf{\prime -}}{\mmf{\prime +}}\\
\omit\rlap{$\Bp{46} = -\frac{\textstyle-\mb \mpp (\Epsi + \mpp) + (\Epsi -\Epp + \mpp) n(u,t) + \Epp \mb \sqrt{u}}{\textstyle 4 \mb^2 \mpp \sqrt{u - u_-}}\sqrt{\Epp + \mpp}\frac{\textstyle\mmf{\prime -}}{\textstyle\mmf{\prime +}}$}\\
\Bp{51} &= \frac{\mpp \sqrt{\Epp + \mpp} (\sqrt{u}-\mpp) \sqrt{u - u_-}}{ 4 (\Eb + \mb) u} &
\Bp{53} &= \frac{\sqrt{\Epp + \mpp} n(u,t) (\sqrt{u}-\mpp) \sqrt{u - u_-}}{4 \mb (\Eb + \mb) \sqrt{u}} \\
\Bp{61} &= -\frac{\mpp \sqrt{\Epp + \mpp} \sqrt{u - u_-}}{4 (\Eb + \mb) u} &
\Bp{63} &= - \frac{\sqrt{\Epp + \mpp} n(u,t) \sqrt{u - u_-}}{4 \mb (\Eb + \mb) \sqrt{u}},
\end{align*}

\begin{align*}
\Pp{26} &= -\frac{\mb \mpsi^2 - \Eb (2 \Epsi \mpp - \mpp^2 + s)}{4 \mb^2 \mpp \sqrt{\Epp + \mpp} }\sqrt{u - u_-}\frac{\mmf{\prime -}}{\mmf{\prime +}}&
\Pp{46} &=  \frac{\Eb \Epsi \sqrt{u - u_-}}{4 \mb^2 \mpp \sqrt{\Epp + \mpp}}\frac{\mmf{\prime -}}{\mmf{\prime +}}\\
\Pp{54} &= \frac{(\mpp + \sqrt{u}) \sqrt{u - u_-}}{4 \mb \sqrt{\Epp + \mpp}}\frac{\mmf{\prime -}}{\mmf{\prime +}}&
\Pp{56} &= -\frac{n(u,t) (\mpp + \sqrt{u})) \sqrt{u}}{ 4 \mb^2 \mpp \sqrt{\Epp + \mpp} }\sqrt{u - u_-}\frac{\mmf{\prime -}}{\mmf{\prime +}}\\
\Pp{64} &= \frac{\sqrt{u - u_-}}{4 \mb \sqrt{\Epp + \mpp} }\frac{\mmf{\prime -}}{\mmf{\prime +}}&
\Pp{66} &= -\frac{n(u,t) \sqrt{u} \sqrt{u - u_-}}{4 \mb^2 \mpp \sqrt{\Epp + \mpp}}\frac{\mmf{\prime -}}{\mmf{\prime +}},
\end{align*}
and
\begin{align*}
 \Rp{13} &= -\frac{(\mpp + \sqrt{u}) \sqrt{u - u_-}}{ 2 \mb (\Eb + \mb) \sqrt{\Epp + \mpp} \sqrt{u}}&
\Rp{16} &=-\frac{(u -\mpp \sqrt{u}) \sqrt{\Epp + \mpp} }{ 2 \mb^2 \mpp \sqrt{u - u_-}}\frac{\mmf{\prime -}}{\mmf{\prime +}}\\
\Rp{23} &= \frac{ n(u,t) + 2 \Eb (\mpp + \sqrt{u})}{4 \mb (\Eb + \mb) \sqrt{\Epp + \mpp} u} \sqrt{u - u_-}&
\Rp{33} &= -\frac{\sqrt{u - u_-}}{2 \mb (\Eb + \mb) \sqrt{\Epp + \mpp} \sqrt{u}}\\
\Rp{36} &= \frac{\sqrt{\Epp + \mpp} \sqrt{u}}{2 \mb^2 \mpp \sqrt{u - u_-}}\frac{\mmf{\prime -}}{\mmf{\prime +}}&
\Rp{43} &= \frac{\Eb\sqrt{u - u_-}}{4 \mb (\Eb + \mb) \sqrt{\Epp + \mpp}  s} \\
\Rp{46} &= -\frac{\Eb \sqrt{\Epp + \mpp}}{4 \mb^2 \mpp \sqrt{u - u_-}}\frac{\mmf{\prime -}}{\mmf{\prime +}}&
\Rp{53} &=\frac{(\mpp + \sqrt{u}) \sqrt{u - u_-}}{4 \mb \sqrt{\Epp + \mpp} \sqrt{u}}\\
\Rp{56} &= -\frac{(\Eb - \mb) (\Epp + \Epsi - \mpp) \sqrt{\Epp + \mpp}}{ 4 \mb^2 \mpp \sqrt{u - u_-} }\frac{\mmf{\prime -}}{\mmf{\prime +}}&
\Rp{63} &= \frac{\sqrt{u - u_-}}{4 \mb \sqrt{\Epp + \mpp} \sqrt{u}}\\
\Rp{66} &= -\frac{(\Eb - \mb) \sqrt{u} \sqrt{\Epp + \mpp}}{4 \mb^2 \mpp  \sqrt{u - u_-}}\frac{\mmf{\prime -}}{\mmf{\prime +}}.
\end{align*}

\section{A practical 
covariant parameterization for the amplitude}
\label{app:practical}
We combine the $s$ and $u$-channel PV and PC isobars. The full covariant amplitude reads~\footnote{For simplicity, we do not specify the helicities explicitly in Eq.~\eqref{eq:total.A}, since matching the helicities in the $s$- and $u$-channel would induce additional Wigner rotations, which eventually cancel when the amplitude is squared and summed over the polarizations. See for example~\cite{Aaij:2015tga,Chen:2017gtx}.}
\begin{align}\label{eq:total.A}
  \mathcal{A}(s,t,u) &= \bar u(p) \sum_{i=1}^{12} \sum_{x=s,u} C_i^{(x)}(x,t)\, M_i^\mu  u(\Lambdab) \epsilon^{*\mu}(\psi),
\end{align}
with $x=s,u$. The tensors $M_{1\cdots 6}$ have been introduced in Eq.~\eqref{eq:cgln}, and we define the $M_{7\cdots 12} \equiv M'_{1\cdots 6}$ for the PV tensors in Eq.~\eqref{eq:cglnpv}. We square the amplitude, contract with the leptonic tensor which describes the $\psi \to \mu^+ \mu^-$ decay, and sum over polarizations
\begin{align}
  \overline{ \sum_\text{pol}}|\mathcal{A}|^2 &= 2 \Bigg[\sum_{i=1}^{12} \sum_{x=s,u}\sum_{i'=1}^{12} \sum_{x'=s,u}  \tr\left[ \left(\slashedpb + \mb\right) M_i^\mu \left(\slashedppp + \mpp\right)  \gamma^0 (M_{i'}^\dagger)^\nu \gamma^0 \right] \nonumber \\
  & \qquad\qquad C^{(x)}_i(x,t)\,C^{(x')*}_{i'} (x',t) \Bigg]\left(l^1_\mu l^2_\nu + l^1_\nu l^2_\mu - g_{\mu\nu} \frac{\mpsi^2}{2}\right),
\end{align}
with $l^1$ and $l^2$ the momenta of $\mu^+$ and $\mu^-$, respectively. This amplitude is has bilinear form in the couplings of the intermediate resonances. We use $R={x,j,\eta,L,S,\eta_b}$ as a collective index, to indicate a resonance in the channel $x=s,u$, having spin $j$, naturality $\eta$, coupling to the $\Lambdab \psi, p \psi$ state in spin $S$ and orbital momentum $L$, and naturality of the \Lambdab $\eta_b$.\footnote{For the $u$-channel resonances $\eta = (-1)^{j+L+1/2}$, for the $s$-channel resonances $\eta = \eta_b(-1)^{j+L+1/2}$, where $\eta_b = 1$ ($-1$) for PC (PV) processes.}
\begin{align}\label{eq:Asq}
  \overline{ \sum_\text{pol}}|\mathcal{A}|^2 &= 2 \sum_{R,R'} g_R(x)\, g^*_{R'}(x')  \tr\left[ \left(\slashedpb + \mb\right) \left(\sum_{i=1}^{12} M_i^\mu \mathcal{C}^{R}_i \right) \left(\slashedppp + \mpp\right)  \gamma^0 \left(\sum_{i'=1}^{12} M_{i'}^\nu \mathcal{C}^{R'}_{i'}\right)^\dagger \gamma^0 \right] \nonumber \\
  & \qquad\qquad \times\left(l^1_\mu l^2_\nu + l^1_\nu l^2_\mu - g_{\mu\nu} \frac{\mpsi^2}{2}\right).
\end{align}

The functions $g_R(x)$ encode all the information about the dynamics, and may be parameterized as Breit-Wigners times the customary Blatt-Weisskopf factors. The scalar $\mathcal{C}^{R}_i$ functions depend only on kinematics
\begin{align}
\begin{pmatrix}
\mathcal{C}^{R}_{1\cdots 6} \\
\mathcal{C}^{R}_{7\cdots 12}
\end{pmatrix}
&= \text{fact}(x) \begin{pmatrix}
\text{Mat}(x) & 0\\
0 & \text{Mat}'(x)
\end{pmatrix}
\begin{pmatrix}
F^{R}_{1\cdots 6}(x,t) \\
F^{R}_{7\cdots 12}(x,t)
\end{pmatrix},
\end{align}
where
\begin{equation}
\text{fact}(x) = \left\{\begin{matrix} \frac{\sqrt{\Eb(s)+\mb}}{\sqrt{\Epp(s) + \mpp}} & \text{for }x=s,\\
\scriptstyle{\sqrt{\Eb(u) + \mb}} & \text{for }x=u,\end{matrix}\right.
\end{equation}
\begin{align}
\Eb(s) &= \frac{s + \mb^2 - \mpsi^2}{2\sqrt{s}},
& \Epsi(s) &= \frac{s - \mb^2 + \mpsi^2}{2\sqrt{s}},
& \Epp(s) &= \frac{s + \mpp^2 - \mK^2}{2\sqrt{s}},\\
p_s &= \frac{\lambda^{1/2}(s,\mb^2,\mpsi^2)}{2\sqrt{s}},
& q_s &= \frac{\lambda^{1/2}(s,\mpp^2,\mK^2)}{2\sqrt{s}}, & &\\
\Eb(u) &= \frac{u + \mb^2 - \mK^2}{2\sqrt{u}},
& \Epsi(u) &= \frac{u - \mpp^2 + \mpsi^2}{2\sqrt{u}},
& \Epp(u) &= \frac{u + \mpp^2 - \mpsi^2}{2\sqrt{u}}, \\
 p_u &= \frac{\lambda^{1/2}(u,\mb^2,\mK^2)}{2\sqrt{u}},
& q_u &= \frac{\lambda^{1/2}(u,\mpp^2,\mpsi^2)}{2\sqrt{u}}, &&
\end{align}
where $\lambda$ is the K\"all\'en triangular function, and
\begin{align*}
\text{Mat}(s)&\text{ is given as $\mathcal{M}^{-1}$ in App.~\ref{app:matrices},} & \text{Mat}'(s)&\text{ is given as $(\mathcal{B}/p_s^2 + \text{Reg})$ in App.~\ref{app:PV},}\\
\text{Mat}(u)&\text{ is given as $(\mathcal{B}/q_u^2 + \mathcal{P}/(u-u_-) + \text{Reg})$ in App.~\ref{app:uchannel},} & \text{Mat}'(u)&\text{ is given as $(\mathcal{B}/q_u^2 + \mathcal{P}/(u-u_-) + \text{Reg})$ in App.~\ref{app:uchannelpv}.}
\end{align*}
The matrices $\text{Mat}^{(\prime)}(x)$ will be available for download on the JPAC website~\cite{JPACweb}.
The functions $F^{R}_{i}$ contain the kinematical dependence of the KSF-PCHAs,
\begin{align}\label{eq:pchavector}
F^{R}_{1\cdots 6}(x,t) &= \begin{pmatrix} F^{R+,\text{PC}}_{+}(x,t) & F^{R+,\text{PC}}_{0}(x,t) & F^{R+,\text{PC}}_{-}(x,t) & F^{R-,\text{PC}}_{+}(x,t) & F^{R-,\text{PC}}_{0}(x,t) & F^{R-,\text{PC}}_{-}(x,t)\end{pmatrix},\\
F^{R}_{7\cdots 12}(x,t) &= \begin{pmatrix} F^{R+,\text{PV}}_{+}(x,t) & F^{R+,\text{PV}}_{0}(x,t) & F^{R+,\text{PV}}_{-}(x,t) & F^{R-,\text{PV}}_{+}(x,t) & F^{R-,\text{PV}}_{0}(x,t) & F^{R-,\text{PV}}_{-}(x,t)  \end{pmatrix},
\end{align}
with
\begin{subequations}
\label{eq:our_result}
\begin{align}
 F^{R\bar \eta, \text{PC}}_{\lambda}(s,t) &= \frac{1}{4\pi} (2j+1)(p_s q_s)^{j-|1/2-\lambda|} \Bigg[ \delta_{\eta,\bar \eta}  \hat d^{j+}_{1/2 - \lambda,1/2}(z_s) \left(\frac{p_s \sqrt{s}}{\mpsi}\right)^{(1 + \eta)\delta_{j,1/2}} + \delta_{\eta,-\bar \eta}  \left(\frac{p_s \mpp}{q_s \mpsi}\right)^{\bar \eta} \hat d^{j-}_{1/2 - \lambda,1/2}(z_s)\Bigg] \nonumber \\
 &\qquad\qquad  \times {\textstyle\left\langle \frac{1}{2}, \frac{1}{2}; 1, -\lambda \,\Big|\,S, \frac{1}{2} - \lambda \right\rangle\left\langle S,\frac{1}{2} - \lambda ; L ,0 \,\Big|\,j,\frac{1}{2} - \lambda \right\rangle} \nonumber \\
 &\qquad \qquad  \times \left(\frac{\Epsi(s)}{\mpsi}\right)^{(1-|\lambda|) (1 - \delta_{j,1/2}\delta_{\eta,1})} \left(\frac{\mpsi \mpp}{\eta \bar{\eta}\sqrt{s}}\right)^{\delta_{\lambda,-1}} (p_s)^{L-j+1 + \eta/2}, \label{eq:shitone}  \\
  F^{R\bar \eta, \text{PV}}_{\lambda}(s,t) &= \frac{1}{4\pi} (2j+1)(p_s q_s)^{j-|1/2-\lambda|} \Bigg[ \delta_{\eta,\bar \eta}  \hat d^{j+}_{1/2 - \lambda,1/2}(z_s)\left(\frac{p_s \sqrt{s}}{\mpsi}\right)^{(1 - \eta)\delta_{j,1/2}}  + \delta_{\eta,-\bar \eta}  \left(\frac{\mpsi \mpp}{s\,p_s q_s }\right)^{\bar \eta} \hat d^{j-}_{1/2 - \lambda,1/2}(z_s)\Bigg] \nonumber \\
 &\qquad\qquad \times {\textstyle\left\langle \frac{1}{2}, \frac{1}{2}; 1, -\lambda \,\Big|\,S, \frac{1}{2} - \lambda \right\rangle\left\langle S,\frac{1}{2} - \lambda ; L ,0 \,\Big|\,j,\frac{1}{2} - \lambda \right\rangle} \nonumber \\ &\qquad \qquad \times \left(\frac{\Epsi(s)}{\mpsi}\right)^{(1-|\lambda|)(1-\delta_{j,1/2}\delta_{\eta,-1})} \left(\frac{\mpsi \mpp}{\eta \bar{\eta}\sqrt{s}}\right)^{\delta_{\lambda,-1}} (p_s)^{L-j+1 - \eta/2},\\
  F^{R\bar \eta, \text{PC}}_{\lambda}(u,t) &= \frac{1}{4\pi} (2j+1)(p_u q_u)^{j-|1/2-\lambda|} \Bigg[ \delta_{\eta,\bar \eta}  \hat d^{j+}_{1/2,1/2 - \lambda}(z_u) \left(\frac{q_u \sqrt{u}}{\mpp}\right)^{(1 + \eta)\delta_{j,1/2}} + \delta_{\eta,-\bar \eta}  \left(\frac{q_u\mb}{p_u\mpp (u - u_-)}\right)^{\bar \eta} \hat d^{j-}_{1/2,1/2 - \lambda}(z_u)\Bigg] \nonumber \\
 &\qquad\qquad \times {\textstyle\left\langle \frac{1}{2}, \frac{1}{2}; 1, -\lambda \,\Big|\,S, \frac{1}{2} - \lambda \right\rangle\left\langle S,\frac{1}{2} - \lambda ; L ,0 \,\Big|\,j,\frac{1}{2} - \lambda \right\rangle} \nonumber\\
 &\qquad\qquad\times\left(\frac{\Epp(u)+\mpp}{2\mpp}\right)^{\frac{1-\eta}{2}}\left(\frac{\mb \mpp}{\eta \bar{\eta}\sqrt{u}}\right)^{\delta_{\lambda,-1}} (q_u)^{L-j+1 + \eta/2},\\
   F^{R\bar \eta, \text{PV}}_{\lambda}(u,t) &= \frac{1}{4\pi} (2j+1)(p_u q_u)^{j-|1/2-\lambda|} \Bigg[ \delta_{\eta,\bar \eta}  \hat d^{j+}_{1/2,1/2 - \lambda}(z_u) \left(\frac{q_u \sqrt{u}}{\mpp}\right)^{(1 + \eta)\delta_{j,1/2}} + \delta_{\eta,-\bar \eta}  \left(\frac{q_u\mb p_u\mpp}{u - u_-}\right)^{\bar \eta} \hat d^{j-}_{1/2,1/2 - \lambda}(z_u)\Bigg] \nonumber \\
 &\qquad\qquad \times {\textstyle\left\langle \frac{1}{2}, \frac{1}{2}; 1, -\lambda \,\Big|\,S, \frac{1}{2} - \lambda \right\rangle\left\langle S,\frac{1}{2} - \lambda ; L ,0 \,\Big|\,j,\frac{1}{2} - \lambda \right\rangle} \nonumber\\
 &\qquad\qquad\times\left(\frac{\Epp(u)+\mpp}{2\mpp}\right)^{\frac{1-\eta}{2}} \left(\frac{\mb \mpp}{\eta \bar{\eta}\sqrt{u}}\right)^{\delta_{\lambda,-1}}  (q_u)^{L-j+1 + \eta/2}.
\end{align}
\end{subequations}
We remind the reader that the KSF-PCHAs contain contributions from partial waves of both naturalities, which we explained when we introduced these amplitudes in Eq.~\eqref{eq:ksf-pcha}. Therefore, $\bar \eta$ is the index of the naturality of the PCHAs, related to the entries of the vector in Eq.~\eqref{eq:pchavector}, and does not coincide with the naturality of the intermediate resonance $\eta$. To make this compact form more understandable, we show the meaning of the factors in $F^{R\bar \eta, \text{PC}}_{\lambda}(s,t)$. Let us consider the example of the $\Lambda(1520)$ with $j^P = \frac{3}{2}^-$ (and naturality $\eta = +$), and coupling to $p \Km$ in $L=2$ and $S=3/2$. The factors in Eq.~\eqref{eq:shitone} are:
\begin{itemize}
\item  $(p_s q_s)^{j-\abs{1/2-\lambda}}$ is the factor $(p_s q_s)^{j-M}$ that cancels the threshold and pseudothreshold singularities of the $\hat d^{j+}_{1/2,1/2-\lambda}(z_s)$. It corresponds to the barrier factors compatible with the minimal $L$ available in a given helicity;
\item  $\delta_{\eta,\bar \eta}  \hat d^{j+}_{1/2,1/2 - \lambda}(z_s)$: since the $\Lambda(1520)$ is natural, it will appear as leading term in the natural $F^{R +, \text{PC}}_{\lambda}(s,t)$;
\item $\left(\frac{p_s \sqrt{s}}{\mpsi}\right)^{(1 + \eta)\delta_{j,1/2}}$ is the special factor appearing for $j^P = \frac{1}{2}^+$. In the case at hand, the factor is $1$; see Eq.~\eqref{eq:Kfactorshalfnat};
\item  $\delta_{\eta,-\bar \eta}  \left(\frac{p_s \mpp}{q_s \mpsi}\right)^{\bar \eta} \hat d^{j-}_{1/2 - \lambda,1/2}(z_s)$: since the $\Lambda(1520)$ is natural, it will appear as subleading term in the $F^{R -, \text{PC}}_{\lambda}(s,t)$, with $\left(\frac{p_s \mpp}{q_s \mpsi}\right)^{-1}$ the mismatch factor between natural and unnatural KSF-PCHAs; see the factor $K^{-\eta}_{MN}/K^{\eta}_{MN}$ Eq.~\eqref{eq:ksf-pcha};
\item  ${\textstyle\left\langle \frac{1}{2}, \frac{1}{2}; 1, -\lambda \,\Big|\,S, \frac{1}{2} - \lambda \right\rangle\left\langle S,\frac{1}{2} - \lambda ; L ,0 \,\Big|\,j,\frac{1}{2} - \lambda \right\rangle}$ are the standard Clebsch-Gordan coefficients that appear in the \ls construction. In our example, $L=2,\,S=3/2$;
\item  $\left(\frac{\Epsi(s)}{\mpsi}\right)^{(1-|\lambda|) (1 - \delta_{j,1/2}\delta_{\eta,1})}$ is the energy-dependent factor derived via our construction. Since the $j^P=\frac{1}{2}+$ case evades the conspiracy equation, there is no need to introduce that factor in that case. See Eq.~\eqref{eq:AzeroEpsiovermpsi};
\item $\left(\frac{\mpsi \mpp}{\eta \bar{\eta}\sqrt{s}}\right)^{\delta_{\lambda,-1}}$ is the mismatch factor between the $\hat A^{j\eta}_{+,+-}(s)$ and $\hat A^{j\eta}_{+,++}(s)$. See Eq.~\eqref{eq:Ahatdefschannel};
\item  $(p_s)^{L-j+1 + \eta/2}= 1$ if $L$ is minimal, or $p_s^2$ if it is nonminimal. In our example $L=2$ is nonminimal (the minimal $L$ allowed is $L=0$) and the factor appears.
\end{itemize}

\bibliographystyle{apsrev4-1}
\bibliography{quattro}
\end{document}